\documentclass[aps, reprint, prl, twocolumn, superscriptaddress,amsmath,amssymb,noeprint]{revtex4-2}

\usepackage{times}
\usepackage{amsmath}
\usepackage{amssymb}
\usepackage{xfrac}
\usepackage{dsfont}
\usepackage{graphicx}
\usepackage{braket}
\usepackage{bbold}
\usepackage[normalem]{ulem}
\usepackage{bm}
\usepackage{bbm}
\usepackage{color}
\usepackage{xcolor}
\usepackage[normalem]{ulem}
\usepackage{pdfpages}
\usepackage{pgffor}
\usepackage{float}

\makeatletter
\AtBeginDocument{\let\LS@rot\@undefined}
\makeatother

\usepackage[colorlinks=true, urlcolor=blue, linkcolor=blue, citecolor=blue]{hyperref}
\usepackage{siunitx}

\newcommand{\Hop}{\hat{H}}
\newcommand{\zop}{\hat{z}}
\newcommand{\pare}[1]{\left( {#1} \right)}
\newcommand{\spare}[1]{\left[ {#1} \right]}

\newcommand{\Dpa}{\Delta_\text{pa}}
\newcommand{\W}{\Omega}
\newcommand{\avg}[1]{\ensuremath{\langle #1 \rangle}}

\newcommand{\berkeleyphy}{Department of Physics, University of California, Berkeley, California 94720}
\newcommand{\CIQC}{Challenge Institute for Quantum Computation, University of California, Berkeley, California 94720}
\newcommand{\LBL}{Materials Sciences Division, Lawrence Berkeley National Laboratory, Berkeley, California 94720}
\newcommand{\columbia}{Department of Physics, Columbia University, New York, NY 10027}
\newcommand{\florida}{Department of Physics, University of South Florida, Tampa, Florida 33620}
\newcommand{\CSIC}{Instituto de F\'isica Fundamental - Consejo Superior de Investigaciones Cient\'ifica (CSIC), Madrid, Espa\~na}

\begin{document}

\title{Optomechanical self-organization in a mesoscopic atom array}


\author{Jacquelyn Ho}
\affiliation{\berkeleyphy}
\affiliation{\CIQC}

\author{Yue-Hui Lu}
\affiliation{\berkeleyphy}
\affiliation{\CIQC}

\author{Tai Xiang}
\affiliation{\berkeleyphy}
\affiliation{\CIQC}

\author{Cosimo C. Rusconi}
\affiliation{\columbia}
\affiliation{\CSIC}

\author{Stuart J. Masson}
\affiliation{\columbia}
\affiliation{\florida}

\author{Ana Asenjo-Garcia}
\affiliation{\columbia}

\author{Zhenjie Yan}
\affiliation{\berkeleyphy}
\affiliation{\CIQC}

\author{Dan M. Stamper-Kurn}
\email[]{dmsk@berkeley.edu}
\affiliation{\berkeleyphy}
\affiliation{\CIQC}
\affiliation{\LBL}
\begin{abstract}

Increasing the number of particles in a system often leads to qualitative changes in its properties, such as breaking of symmetries and the appearance of phase transitions. This renders a macroscopic system fundamentally different from its individual microscopic constituents. Lying between these extremes, mesoscopic systems exhibit microscopic fluctuations that influence behavior on longer length scales, leading to critical phenomena and dynamics. Therefore, tracing the properties of well-controlled mesoscopic systems can help bridge the gap between an exact description of few-body microscopic systems and the emergent description of many-body systems. Here, we explore mesoscopic signatures of an optomechanical self-organization phase transition using arrays of cold atoms inside an optical cavity. By precisely engineering atom-cavity interactions, we reveal how critical behavior depends on atom number, identify characteristic dynamical behaviors in the self-organized regime, and observe a finite optomechanical susceptibility at the critical point. These findings advance our understanding of particle-number- and time-resolved properties of phase transitions in mesoscopic systems.

\end{abstract}

\maketitle

The scale of a physical system plays a key role in determining its material properties. With increasing size comes increasing complexity, and complexity leads to the emergence of macroscopic phenomena that, often, cannot be extrapolated from microscopic behaviors~\cite{andersonMoreDifferent1972}. A well-known example of this paradigm is the appearance of phase transitions in the thermodynamic limit, i.e.\ as a system's particle number tends to infinity. In between the microscopic and macroscopic extremes lies the mesoscopic regime. Mesoscopic systems not only exhibit characteristics of macroscopic systems, but also host a variety of rich properties themselves. In systems of intermediate size, assumptions of local equilibrium break down and internal microstructures may affect properties on longer length scales~\cite{jouMesoscopicTransportEquations2011}. The fact that microscopic fluctuations have a measurable influence on mesoscopic scales also makes mesoscopic systems an interesting playground for the study of critical phenomena and phase transitions~\cite{iachelloQuantumPhaseTransitions2004,vidalFinitesizeScalingExponents2006,stitelySuperradiantSwitchingQuantum2020}.

Technical advances in engineering, probing and measuring mesoscopic systems have propelled experiments in a variety of physical platforms. Notable examples include solid state systems~\cite{karapetrovDirectObservationGeometrical2005, birdQuantumTransportOpen1997}, nanostructured~\cite{knollDirectImagingMesoscale2004} and nanomechanical systems~\cite{mathenyExoticStatesSimple2019,wollackQuantumStatePreparation2022,mirhosseiniSuperconductingQubitOptical2020,renTopologicalPhononTransport2022}, nanomagnets~\cite{andradeDampedPrecessionMagnetization2006,thirionSwitchingMagnetizationNonlinear2003}, and cold atoms and ions~\cite{bayhaObservingEmergenceQuantum2020,brantutConductionUltracoldFermions2012,zeiherTrackingEvaporativeCooling2021, gormanEngineeringVibrationallyAssisted2018,guoObservation2D1D2024,safavi-nainiVerificationManyIonSimulator2018,kaufmanQuantumThermalizationEntanglement2016}. Breakthroughs in the control of atoms trapped in optical
tweezer arrays~\cite{kaufmanQuantumScienceOptical2021,browaeysManybodyPhysicsIndividually2020} represent a leap forward in precisely constructing
mesoscopic, ultracold atomic systems. We show that deterministic control over particle number and interaction strength can be achieved with an optical tweezer array inside a standing-wave optical cavity. This system uniquely lends itself to the detailed study of optomechanics in mesoscopic samples. 

We focus on the mechanical self-organization generated by light-mediated forces among optically excited atoms within an optical cavity~\cite{ritschColdAtomsCavitygenerated2013,mivehvarCavityQEDQuantum2021}. When such atoms are driven by coherent light above a certain intensity, they exhibit a phase transition in which they spontaneously scatter light into the cavity. That light creates a spatially dependent potential on the atoms that causes them to self-organize and superradiantly emit into the cavity with a phase that breaks $\mathbb{Z}_2$ symmetry. This self-organization phase transition maps onto the Dicke model~\cite{kirtonIntroductionDickeModel2019,baumannDickeQuantumPhase2010}. Previous experiments on the optomechanical Dicke phase transition \cite{baumannDickeQuantumPhase2010} and other phenomena arising from cavity-mediated forces among atoms \cite{leonardSupersolidFormationQuantum2017, moralesCouplingTwoOrder2018,kongkhambutObservationContinuousTime2022,zhiqiangNonequilibriumPhaseTransition2017,blackObservationCollectiveFriction2003} were conducted on bulk macroscopic samples, with particle numbers $N \gtrsim 10^5$.  Here, we realize the mesoscopic limit of the Dicke phase transition in deterministically prepared tweezer arrays containing between 10 and 22 atoms. We observe a critical point that depends on atom number, a finite switching time scale between self-organized states, and a collectively altered susceptibility to external forces. This setup allows us to study the crossover from the few- to many-body regime of phase transitions, advancing our knowledge of finite size effects and dynamical behavior in mesoscopic systems.  

\section{Cavity-Mediated interactions and Self-organization}
\begin{figure*}
    \centering
    \includegraphics{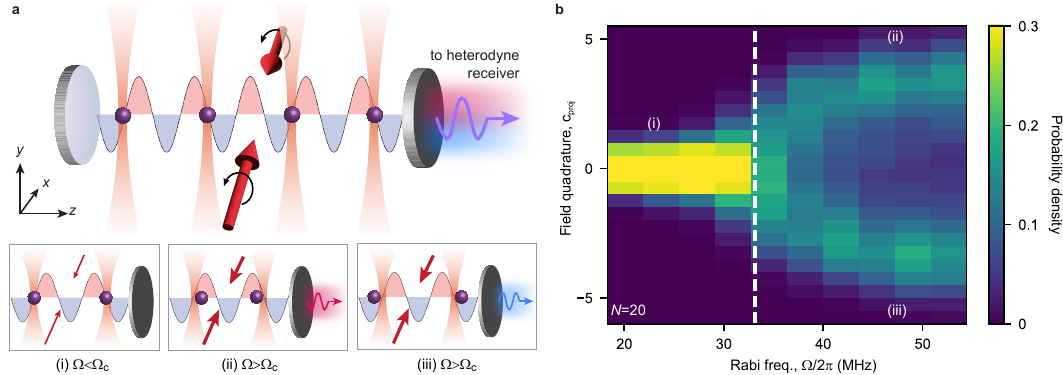}
    \caption{\textbf{Cavity-induced self-organization.} \textbf{a,} Schematic of the experimental setup. Optically tweezed atoms are placed at the nodes of the cavity field at half-integer wavelength spacing (1.5\,$\lambda$ shown in figure for convenience). The atoms are driven by a pair of circularly polarized counter-propagating pump beams. The transmitted cavity field is detected with a heterodyne receiver. Insets (i-iii) illustrate self-organization. (i) At pump strengths below the critical point, the atoms remain centered at the cavity nodes and on average emit zero field. Above the critical point, atoms self-organize by moving toward either the (ii) positive (red) or (iii) negative (blue) antinodes. \textbf{b,} Probability distributions of cavity field projections from repeated measurements. The results shown are based on data collected with $N=20$, $\mathrm{\Delta_{pc}}=-2\pi\times 1.9\,\mathrm{MHz}$ and $\mathrm{\Delta_{pa}}=-2\pi\times 80\,\mathrm{MHz}$. The field distributions are plotted for different values of $\mathrm{\Omega}$. The data are averaged over $5\,\mu\mathrm{s}$ intervals and are sampled from the first $100\,\mu\mathrm{s}$ after the pump ramp-up has finished. The labels (i), (ii), and (iii) correspond to the drawings in (a). The white dashed line indicates the critical point fitted from the data.
    }
    \label{fig:M1}
\end{figure*} 

To realize mesoscopic self-organization, we use the experimental setup schematically shown in Fig.~\ref{fig:M1}a. It contains a one-dimensional tweezer array into which we load a deterministic number $N$ of $^{87}$Rb atoms. The array is aligned along the axis of a high-finesse Fabry-P\'{e}rot optical cavity (the $z$-axis). The cavity is placed in-vacuum and has a TEM$_{00}$ mode whose frequency, $\omega_\mathrm{c}$, lies near the resonance frequency, $\omega_\mathrm{a}$, of the $^{87}$Rb $F=2 \rightarrow F^\prime=3$ $D_2$ transition (wavelength $\lambda = 780\,$nm)~\cite{deistMidCircuitCavityMeasurement2022, deistSuperresolutionMicroscopyOptical2022}. The atoms, prepared in the $F=2$ ground state manifold, are exposed to a circularly polarized standing wave pump beam of frequency $\omega_\mathrm{p}$ (detuned from the atoms by $\mathrm{\Delta_{pa}}=\omega_\mathrm{p}-\omega_\mathrm{a}$ and from the cavity by $\mathrm{\Delta_{pc}}=\omega_\mathrm{p}-\omega_\mathrm{c}$) propagating along the $x$-axis. The circularly polarized pump creates a fictitious magnetic field along $x$ through vector light shifts and defines the atomic quantization axis. The pump drives the $\sigma^-$ electric dipole transition of the atoms and its power is linearly ramped up to the desired value in $50\,\mathrm{\mu s}$. The initial ramp optically pumps the atoms into the $\lvert F=2, m_F=-2\rangle$ stretched state, whence the atoms are excited on the cycling transition, thereby ensuring that no dynamics among the $\lvert F=2,m_F\rangle$ states interfere with the location of the critical point for self-organization. We achieve a maximal vacuum Rabi coupling of $g_0 = 2 \pi \times 3.1/\sqrt{2}$\,MHz between the $y$-polarized cavity field and the atomic $\sigma^-$ transition. The cavity half-linewidth is $\kappa = 2 \pi \times 0.53\,\mathrm{MHz}$, resulting in a sub-$\mu$s cavity response time defined by $1/\kappa$. Given that the atomic half-linewidth is $\gamma = 2 \pi \times 3.0\,\mathrm{MHz}$, we achieve a single-atom cooperativity
$C = g_0^2/(2 \kappa \gamma) = 1.5$~\cite{kimbleStrongInteractionsSingle1998}. Because the atom-cavity interaction depends strongly on the spatial structure of the pump and cavity standing waves~\cite{yanSuperradiantSubradiantCavity2023a, reimannCavityModifiedCollectiveRayleigh2015, begleyOptimizedMultiIonCavity2016,neuznerInterferenceDynamicsLight2016}, the tweezers are aligned to be all within the same pump antinode wavefront (so the atoms all experience the same pump Rabi frequency, $\mathrm{\Omega}$) and are positioned at cavity nodes spaced $4.5\,\lambda$ apart. By spacing the tweezers by a half-integer number of wavelengths, we eliminate biasing in our system due to center-of-mass drift of the tweezer array with respect to the cavity standing wave. 

Light inside the cavity creates a potential energy landscape for the atoms through the exchange of cavity photons~\cite{mivehvarCavityQEDQuantum2021}. For atom $n$ at position $\hat{z}_n$, the cavity interaction potential, produced by interference between the pump and cavity fields, is proportional to $S\sin(k\hat{z}_n)(\hat{c}+\hat{c}^\dagger)$, where $\hat{c}$ ($\hat{c}^\dagger$) is the bosonic annihilation (creation) operator for a cavity photon, $S\simeq \mathrm{\Omega}g_0/(2\mathrm{\Delta_{pa}})$ is the coherent scattering rate of a pump photon into the cavity, and $k=2\pi/\lambda$. The strength of the potential thus depends on the atom's position inside the cavity and on the magnitude of the cavity field. When the pump is sufficiently detuned from the atomic transition and red-detuned from cavity resonance, the cavity field follows the atomic motion such that $\hat{c}+\hat{c}^\dagger \propto -S \sum_{n=1}^N \sin(k\hat{z}_n)$. In this adiabatic regime, the effective potential is minimized when $\langle \sin(k\hat{z}_n)\rangle =\pm 1$ for all $n$. That is, the most favorable configurations of the system are the ones where the atoms self-organize by collectively moving to cavity-field antinodes with the same sign.

The steady state of the optomechanical system is governed by competition between the cavity-mediated interaction and the tweezer potential, which, for tweezers centered at the cavity nodes, is minimized when $\langle \sin(k\hat{z}_n)\rangle =0$ for all $n$.
This competition gives rise to the emergence of self-organization as a function of $S$, which we tune through $\mathrm{\Omega}$, thus defining a critical pump strength $\mathrm{\Omega_c}$. 
For $\mathrm{\Omega}<\mathrm{\Omega_c}$, the tweezers dominate over the cavity field and the atoms remain centered at the cavity nodes (as shown in Fig.~\ref{fig:M1}a(i)). Although the atom array is uncoupled from the cavity when positioned exactly at the nodes, zero-point and thermal fluctuations shift the atom positions, causing the array to emit light incoherently into the cavity.
For $\mathrm{\Omega}>\mathrm{\Omega_c}$, the negative curvature of the cavity potential overcomes the positive tweezer trap curvature and it becomes energetically favorable for all atoms to move towards antinodes with the same sign and for their emission to interfere constructively. The superradiant field's phase is determined by the sign of the antinodes chosen by the atoms (Fig.~\ref{fig:M1}a(ii-iii)).
For atoms spaced by a half-integer number of wavelengths, adjacent atoms move in opposite directions to achieve self-organization (antiphase motion).
We refer to the collective mode that is linearly coupled to the cavity as the dominant mode.

To confirm the self-organization experimentally, we detect the cavity field $\hat{c}$ and observe $\mathbb{Z}_2$ symmetry breaking. Assuming $\mathrm{\Delta_{pa}}<0$ (the case for all data shown), light scattered by the atoms affects one quadrature of the cavity field,
$c_\mathrm{proj}=\mathrm{Re}(\langle\hat{c}\rangle)\cos\theta+\mathrm{Im}(\langle\hat{c}\rangle)\sin\theta$, where $\theta=\arctan(-\kappa/\mathrm{\avg{\widetilde{\Delta}_{pc}}})$, defined with respect to the phase of the pump field. One finds 

\begin{align}
    \label{c_quad}     c_\mathrm{proj} &=\frac{N\mathrm{\Omega}g_0}{2|\mathrm{\Delta_{pa}}|}\Bigg\langle\frac{\mathrm{\hat{\Theta}}}{\sqrt{\mathrm{\widetilde{\Delta}_{pc}}^2+\kappa^2}}\Bigg\rangle.
\end{align}
We define $\mathrm{\widetilde{\Delta}_{pc}}=\mathrm{\Delta_{pc}}-\sum_{n=1}^{N}g_0^2\sin^2(k\zop_n)/\mathrm{\Delta_{pa}}$ as the effective cavity detuning including the dispersive shift from the atoms. The cavity field is proportional to the operator
\begin{equation}
	\hat{\Theta} \equiv \frac{1}{N}\sum_{n=1}^N  \sin(k \zop_n),
\end{equation}
whose mean value can be interpreted as the order parameter. For atoms trapped at the nodes $\avg{\hat{\Theta}}=0$, while for atoms at the antinodes $\avg{\hat{\Theta}}=\pm 1$. Thus $\avg{\hat{\Theta}}$ takes the role of an effective magnetization for the system~\cite{ritschColdAtomsCavitygenerated2013,Schuetz2015}. 

To detect the complex cavity field $\hat{c}$, we send $y$-polarized emission from one cavity mirror to a heterodyne receiver (see Extended Data Fig.~\ref{fig:S1}) and record the signal for a total time of $250\,\mu \mathrm{s}$, including the $50\,\mu$s pump ramp-up time. We filter the time trace with an adjustable averaging time, ranging from $5\,\mu\mathrm{s}$ to $50\,\mu\mathrm{s}$; we average for a minimum of $5\,\mu$s because increased noise at shorter averaging times compromises the signal-to-noise ratio of the extracted cavity field.
The values obtained at each step of the filter are treated as individual data points for $\hat{c}$. We determine the angle $\theta$ by referencing the phase of the distribution of $\hat{c}$ to that obtained with all atoms placed at antinodes of the same sign. We then project each value of $\hat{c}$ onto the axis defined by $\theta$ to obtain $c_\mathrm{proj}$ (see Methods and Extended Data Fig.~\ref{fig:S2}). Our measured $c_\mathrm{proj}$ is scaled such that $c_\mathrm{proj}^2$ is the average photon number in the cavity. 

To observe bifurcation in the distributions of $c_\mathrm{proj}$, we repeat the measurement many times at different pump strengths and analyze the heterodyne data from $t=0$ to $t=100\,\mu$s (where $t=0$ is the end of the pump ramp-up) using a $5\,\mu$s averaging time. Due to heating of the atoms from continuous exposure to pump light~\cite{arnoldSelfOrganizationThresholdScaling2012,niedenzuKineticTheoryCavity2011}, we constrain this analysis to the $100\,\mu$s span described above. The distributions of $c_\mathrm{proj}$ show a striking bifurcation as the pump strength is increased. An example is shown in Fig.~\ref{fig:M1}b. 
We see that once $\mathrm{\Omega}$ surpasses a threshold near $2\pi\times 30\,\mathrm{MHz}$, the single peak at $c_\mathrm{proj}=0$ splits into two symmetric branches with finite field amplitude. 

The simplest model that captures this physics is purely mechanical and lossless, with an effective Hamiltonian
\begin{equation}\label{eq:Heff}
\begin{split}
	\Hop_\text{eff} \equiv& \sum_{n=1}^N \spare{\frac{\hat{p}_n^2}{2M}+\frac{M}{2}\nu^2\pare{\zop_n-z_{0n}}^2} + \hbar  \mathcal{D} N^2 \hat{\Theta}^2.
\end{split}
\end{equation}
Here, $M$ is the atomic mass of $^{87}$Rb, $\nu=2\pi\times 93\,\mathrm{kHz}$ is the tweezer trap frequency (in the harmonic approximation), $\hbar$ is the reduced Planck constant, and $\mathcal{D}\equiv \pare{\frac{g_0\W}{2\Dpa}}^2 \tilde{\Delta}_\text{pc}/(\tilde{\Delta}_\text{pc}^2+\kappa^2)$ is the strength of the cavity-induced mechanical potential, whose sign depends on $\tilde{\Delta}_\text{pc}$. The tweezer positions are $z_{0n}=4.5\lambda \times n$.

\begin{figure}
    \centering
    \includegraphics{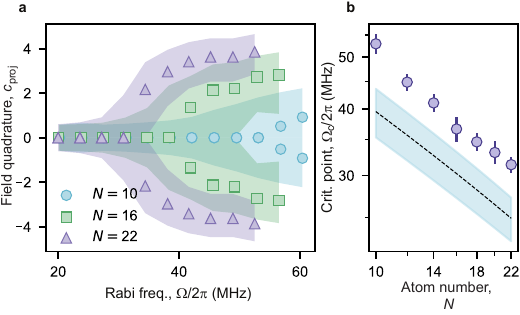}
    \caption{\textbf{Dependence of the critical point on atom number.} \textbf{a,} Bifurcation data for $N=10$ (circles), $N=16$ (squares), and $N=22$ (triangles). Shaded areas show the width of the distribution at half maximum while markers show the maxima of the fitted Boltzmann distrbution. \textbf{b,} 
    Experimentally measured $\mathrm{\Omega_c}$ (circles) for different $N$ plotted on a logarithmic scale and the theoretical prediction of Equation~(\ref{Omega_c}), based on an independent temperature measurement of $T=35\pm10\,\mu$K (black dashed line and light blue shaded area). Error bars represent standard deviations resulting from statistical errors on fits of the parameter $B$ and drift of the pump power over the course of the experiment, estimated to be 10\%. The reported values of $\Omega_\mathrm{c}$ account for systematic errors due to Gaussian detector noise and the Gaussian intensity profile of the pump beam. We note that there is an additional systematic error of +6.0\% and -8.5\% on our calibration of $\Omega$ (see Methods for details). All data in figure are taken at $\mathrm{\Delta_{pa}}=-2\pi\times80\,\mathrm{MHz}$ and $\mathrm{\Delta_{pc}}=-2\pi\times 1.9\,\mathrm{MHz}$. The value of $\mathrm{\Delta_{pa}}$ was chosen to avoid strong atom number-dependence of the atom-induced dispersive shift on the cavity resonance frequency. We also chose the bare cavity resonance frequency to be sufficiently larger than the pump frequency to avoid instabilities arising from the dispersive shift bringing the pump blue-detuned of the cavity. }
    \label{fig:M2}
\end{figure}

However, thermal fluctuations as well as motion of the atoms perpendicular to the cavity axis are relevant for a quantitative description of the experiment. By means of a more refined optomechanical model (see Supplementary Information Section II), the critical pump strength is found to be
\begin{equation}\label{Omega_c}
    \mathrm{\Omega_c}(T)=\sqrt{\frac{2M\nu^2\mathrm{\Delta^2_{pa}}(\mathrm{\Delta^2_{pc}}(T)+\kappa^2)}{Ng^2_0 k^2|\mathrm{\Delta_{pc}}(T)|\hbar \varepsilon(T)}},   
\end{equation}
where $T$ is the atomic motional temperature and the temperature-dependent factor $\varepsilon(T)$ is defined in Supplementary Equation (S29). The dependencies of $\mathrm{\Omega_c}$ on $N$ and $\mathrm{\Delta_{pa}}$ are studied in this work (see Fig.~\ref{fig:M2} and Extended Data Fig.~\ref{fig:S3}). Equation~(\ref{Omega_c}) includes two different thermal effects on the critical pump strength. First, taking an average over the thermal motional shift of the atoms yields a net shift of the cavity resonance $\Delta_\text{pc}(T) \equiv \avg{\tilde{\Delta}_\text{pc}}$. Second, thermal fluctuations allow the atoms to probe the non-linearity of the effective mechanical potential in $\hat{\Theta}^2$. This leads to an increase in the critical pump strength proportional to the factor $1/\varepsilon(T)$ in Equation~(\ref{Omega_c}). Using a temperature of $35\pm10\,\mu$K, determined from an independent measurement~\cite{yanSuperradiantSubradiantCavity2023a}, Equation~(\ref{Omega_c}) predicts $\mathrm{\Omega_c}=2\pi\times (26.5\pm2.5)\,$MHz for the parameter settings in Fig.~\ref{fig:M1}b.

To determine $\mathrm{\Omega_c}$ quantitatively from the data, we draw inspiration from Landau's theory of phase transitions~\cite{landauTHEORYPHASETRANSITIONS1965}. If we only consider the contribution of the dominant mode to the Hamiltonian in Equation~(\ref{eq:Heff}), the potential energy becomes $\frac{1}{2}NM\nu^2\hat{z}_\mathrm{dom}^2+\hbar \mathcal{D}N^2\sin^2 (k\hat{z}_\mathrm{dom})$, where $\hat{z}_\mathrm{dom}$ is the displacement of the dominant mode with respect to the cavity nodes. If we consider the distribution of positions to be a Boltzmann distribution determined by the local potential energy, and expand the potential energy to fourth order in $\hat{z}_\mathrm{dom}$, we obtain a probability distribution $p(z_\mathrm{dom})$ of the form
\begin{equation}
    \label{z_dist}
    p(z_\mathrm{dom})\propto \exp(-Bz_\mathrm{dom}^2-Dz_\mathrm{dom}^4),
\end{equation}
where $B$ and $D$ are parameter-dependent constants and $z_\mathrm{dom}\equiv \langle \hat{z}_\mathrm{dom} \rangle$. According to Landau's theory, a second order phase transition occurs at $B=0$. Because $c_\mathrm{proj}$ is proportional to $z_\mathrm{dom}$ to leading order, we fit the observed distributions of $c_\mathrm{proj}$ to Equation~(\ref{z_dist}) and extract the coefficient $B$ to determine where $\mathrm{\Omega_c}$ occurs.  Analyzing the distributions of $c_\mathrm{proj}$, we observe that the fits to the Boltzmann distribution match the data well (see Extended Data Fig.~\ref{fig:S4} and Methods) even though the system is driven and dissipative. From the data in Fig.~\ref{fig:M1}b, we extract an experimental $\mathrm{\Omega_c}=2\pi\times(33.1^{+2.0_\mathrm{syst}}_{-2.8_\mathrm{syst}}\pm1.6_\mathrm{stat})\,\mathrm{MHz}$.
This measurement is slightly above the prediction of Equation~(\ref{Omega_c}) for our independent measurement of the initial temperature of the trapped atoms, though within error. The slight discrepancy may be caused by heating of the trapped atoms during the experiment, owing to light scattering. In future work, it may be possible to monitor the atoms' temperature during the experiment by observing sideband asymmetry in the heterodyne power spectrum.

\section{Mesoscopic signatures of a phase transition}

We now examine how critical behavior varies with atom number. We first examine the dependence of $\mathrm{\Omega_c}$ on $N$, and find that $\Omega_c$ diminishes as $N$ increases. Quantitatively, for an averaging time of $5\,\mu\mathrm{s}$, we almost recover the $\sim1/\sqrt{N}$ dependence predicted by Equation~(\ref{Omega_c}), shown in Fig.~\ref{fig:M2}b. Though the critical point's scaling with $N$ has been measured before in macroscopic bosonic and fermionic realizations of the superradiant phase transition~\cite{arnoldSelfOrganizationThresholdScaling2012,zhangObservationSuperradiantQuantum2021}, our work shows that the $N$-dependence is also observable in a regime far from the thermodynamic limit. In Fig.~\ref{fig:M2}b, the data appear above the prediction of Equation~(\ref{Omega_c}) if we use the temperature from our independent measurement, implying that the atoms have been heated to a temperature close to $66\,\mu$K at the time of observation.

\begin{figure}
    \centering
    \includegraphics{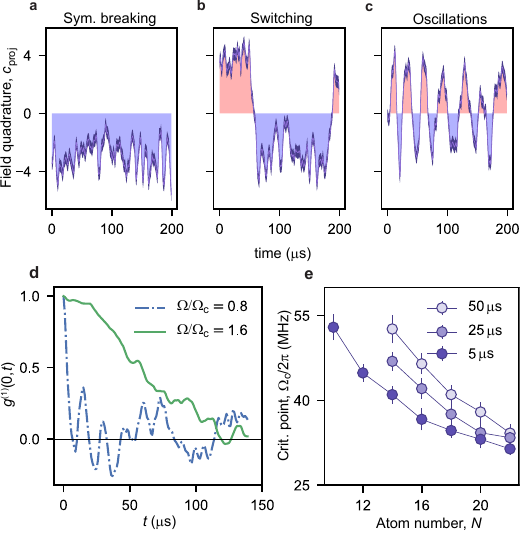}
    \caption{\textbf{Lifetime of the self-organized state.} Above the critical point, we observe three dynamical behaviors in the cavity field time traces: \textbf{a,} symmetry breaking, \textbf{b,} switching, and \textbf{c,} rapid oscillations. Time is measured from the end of the pump ramp-up. Shaded areas show the time spent in the $c_\mathrm{proj}>0$ state (red) and the $c_\mathrm{proj}<0$ state (blue). Shaded purple regions indicate the shot noise level. Data are shown for $N=20,\,\mathrm{\Omega}=2\pi\times45.4\,\mathrm{MHz}$. \textbf{d,} $g^{(1)}(0,t)$ for $N=20$ at  $\mathrm{\Omega}<\mathrm{\Omega_c}$ and $\mathrm{\Omega}>\mathrm{\Omega_c}$. When the system is self-organized, $g^{(1)}(0,t)$ decays on a much longer timescale. \textbf{e,} Dependence of the critical point on $N$ for different averaging times. At larger $N$, the curves converge, indicating that self-organization persists on longer timescales. At lower $N$, the separation between the curves grows due to the transience of the mesoscopic system. The measurements at $25\,\mu$s and $50\,\mu$s for $N<14$ are absent because the transition was not reached in the range of $\mathrm{\Omega}$ values scanned. The plotted $\Omega_\mathrm{c}$ data are subject to the same systematic errors as described in Fig.~\ref{fig:M2}b. Lines are guides to the eye. All data in figure are taken at $\mathrm{\Delta_{pa}}=-2\pi\times 80\,\mathrm{MHz}$ and $\mathrm{\Delta_{pc}}=-2\pi\times1.9\,\mathrm{MHz}$. }
    \label{fig:M3}
\end{figure}

Atomic self-organization in a cavity can be compared with the crossover between a magnetized and unmagnetized state in a spin system~\cite{Schuetz2015}. The behavior of $\avg{\hat{\Theta}}$ above the transition is analogous to that of a
superparamagnet: As discussed in seminal work by N\'eel\,\cite{neelThermoremanentMagnetizationFine1953}, small ferromagnetic particles are subject to thermal fluctuations that cause their magnetization to relax on a time scale that grows with the particle volume. Averaging the magnetization over times much longer than the relaxation time results in zero net magnetization, while averaging over times much shorter than the relaxation time allows for observation of quasistatic magnetic ordering~\cite{knobelSuperparamagnetismOtherMagnetic2008}. We similarly find that mesoscopic self-organization of atoms has a characteristic lifetime that depends on system size. 

To reveal the dynamics of the magnetization order parameter  $\avg{\hat{\mathrm{\Theta}}}$, we take advantage of the cavity's $\mu$s-scale readout capability~\cite{deistMidCircuitCavityMeasurement2022}. Fig.~\ref{fig:M3}a-c show single-shot time traces representative of observed dynamical behaviors in the self-organized regime: symmetry breaking, switching, and rapid oscillations. For a macroscopic atom number, only symmetry breaking should occur, while for a mesoscopic atom number, fluctuations away from equilibrium can lead to switching between stable states of the system~\cite{Schuetz2015}. The rapid oscillation events are rare, occurring in only 2 out of the 180 shots obtained at the shown parameter setting; we expect they result from the energy in the dominant mode being above the energy gap separating the two symmetry-broken states. 

To highlight the difference in dynamical behavior below and above the critical point, we calculate the first-order coherence function of the field quadrature $g^{(1)}(t_1,t_2)$, defined as $g^{(1)}(t_1,t_2)=\langle c^*_\mathrm{proj}(t_1)c_\mathrm{proj}(t_2)\rangle/\sqrt{\langle |c_\mathrm{proj}(t_1)|^2\rangle \langle |c_\mathrm{proj}(t_2)|^2\rangle}$, where the angle brackets denote averaging over multiple shots. In particular, $g^{(1)}(0,t)$ indicates how correlated the field at time $t_2=t$ is to the field at time $t_1=0$, where $t_1=0$ is chosen to be at the end of the pump ramp-up. Symmetry breaking is indicated by $g^{(1)}(0,t)$ remaining high for extended $t$. In Fig.~\ref{fig:M3}d, we see that when $\mathrm{\Omega}/\mathrm{\Omega_c}=0.8$, $g^{(1)}(0,t)$ decays rapidly, indicating the absence of symmetry breaking, and exhibits oscillations at approximately $60\,\mathrm{kHz}$, corresponding to the softened trap frequency. At $\mathrm{\Omega}/\mathrm{\Omega_c}=1.6$, $g^{(1)}(0,t)$ decays on a much longer timescale and oscillations are absent, indicating a prolonged lifetime of the symmetry-broken states. While we are able to observe these qualitative differences in the decay of $g^{(1)}(0,t)$ for $\Omega<\Omega_\mathrm{c}$ and $\Omega>\Omega_\mathrm{c}$, we note that using $g^{(1)}(0,t)$ to determine the location of $\Omega_\mathrm{c}$ does not appear to be a robust method.

We next observe the $N$-dependence of the lifetime of the self-organized state. As discussed above, the characteristic behavior of the system depends on the time over which $\avg{\hat{\mathrm{\Theta}}}$ is averaged. On short time scales, the system displays signatures of symmetry breaking, while on long time scales, the system switches between the two stable states. As a result, in Fig.~\ref{fig:M3}e, we see that the nominal $\mathrm{\Omega_c}$ increases as we increase the averaging time from 5 to $50\,\mu\mathrm{s}$. Remarkably, we see that $\mathrm{\Omega_c}$ increases significantly at lower $N$ but less so at higher $N$. This indicates that as the system size decreases, the characteristic time scale over which the system remains self-organized also decreases. We thus observe a key feature of this mesoscopic system: Averaging $\langle\hat{\mathrm{\Theta}}\rangle$ over times longer than the $N$-dependent lifetime erases signatures of symmetry breaking.

\begin{figure}
    \centering
    \includegraphics{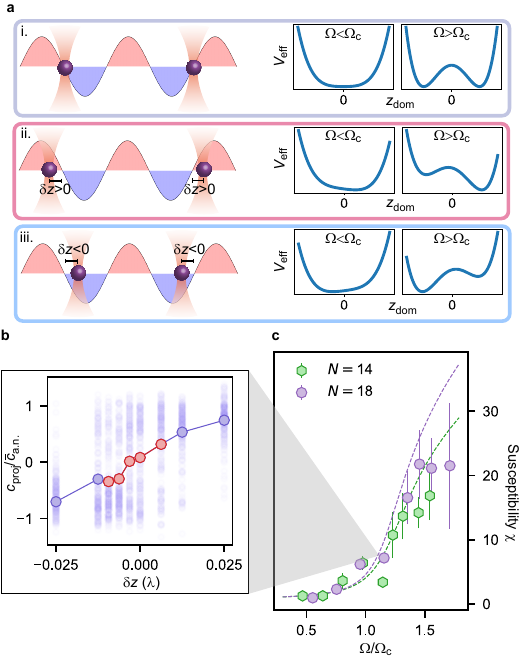}
    \caption{\textbf{Optomechanical susceptibility.} \textbf{a,} Schematic of the susceptibility measurement. The dominant mode of the array ($z_\mathrm{dom}$) is displaced by $\delta z$. When $\delta z=0$, the effective potential $V_\mathrm{eff}$ on the atoms is symmetric about $z_\mathrm{dom}=0$ (i). For $|\delta z|>0$, $V_\mathrm{eff}$ tilts in the direction of the bias (ii-iii). \textbf{b,} Distributions of $c_\mathrm{proj}/\bar{c}_\mathrm{a.n.}$ (small purple points) and their mean values (circles) at various $\delta z$ for $N=18,\,\mathrm{\Omega/\Omega_c}=1.16$. Measurement of $\partial  (\bar{c}_\mathrm{proj}/ \bar{c}_\mathrm{a.n.})/\partial (\delta z)\big\rvert_{\delta z=0}$ is taken by fitting the slope of $\bar{c}_\mathrm{proj}/\bar{c}_\mathrm{a.n.}$ for $-0.01\,\lambda\leq\delta z\leq 0.01\,\lambda$ (red circles). \textbf{c,} Susceptibility $\chi$ at various $\mathrm{\Omega}/\mathrm{\Omega_c}$ for $N=14$ and $N=18$. The data (hexagons and circles) are plotted together with the theoretical prediction based on the Boltzmann distribution for $T=35\,\mu$K (dashed lines). All data in figure are taken at $\mathrm{\Delta_{pa}}=-2\pi\times80\,\mathrm{MHz}$ and $\mathrm{\Delta_{pc}}=-2\pi\times1.9\,\mathrm{MHz}$.}
    \label{fig:M4}
\end{figure}

\section{Finite susceptibility at criticality} 
The positional control afforded by the tweezers not only allows us to observe spontaneous symmetry breaking, but also to explicitly break symmetry with nanometer-scale precision and probe the system's response. By displacing adjacent tweezers in opposite directions by an amount $\delta z$, we introduce a constant force proportional to $\delta z$ on the atoms that biases the dominant mode (Fig.~\ref{fig:M4}a). Generally, the amount a system moves in response to an applied force is characterized by its mechanical susceptibility. In our case, $c_\mathrm{proj}$ serves as a proxy for the atom positions. As such, we obtain the optomechanical susceptibility, $\chi$, which we define as
\begin{equation}
    \label{chi}
    \chi\equiv\frac{1}{k\bar{c}_\mathrm{a.n.}}\frac{\partial \bar{c}_\mathrm{proj}}{\partial (\delta z)}\bigg\rvert_{\delta z=0}.
\end{equation}
Here, $\bar{c}_\mathrm{a.n.}$ is the measurement of $\bar{c}_\mathrm{proj}$ when the tweezers are positioned at the positive antinodes. The overbar indicates an average over the distribution of $c_\mathrm{proj}$ obtained from the first $50\,\mu\mathrm{s}$ of multiple shots. The quantity $\partial  (\bar{c}_\mathrm{proj}/ \bar{c}_\mathrm{a.n.})/\partial (\delta z)\big\rvert_{\delta z=0}$ is measured by fitting the slope of $\bar{c}_\mathrm{proj}/\bar{c}_\mathrm{a.n.}$ in the range $-0.01\,\lambda\leq\delta z\leq 0.01\,\lambda$ for $N=18$ (red circles in Fig.~\ref{fig:M4}b) and $N=14$. We note that the atoms are initialized with root-mean-square thermal fluctuations in position of about 100\,nm, which is comparable to $k^{-1}$, leading to a thermal Lamb-Dicke parameter of about 0.8 (see Supplementary Equation (S18)).

From measurement of $\chi$ for $N=14$ and $N=18$ at different values of $\mathrm{\Omega}/\mathrm{\Omega_c}$ (Fig.~\ref{fig:M4}c), we make two observations: 1) while $\chi$ increases with $\mathrm{\Omega/\Omega_c}$, it does not diverge at the critical point as one would expect for a second-order phase transition in a macrosopic system, and 2) the value of $\chi$ grows with atom number for $\mathrm{\Omega/\Omega_c}>1$. These observations can be explained by a model assuming $\chi\simeq \partial \langle z_\mathrm{dom}\rangle/\partial (\delta z)\big\rvert_{\delta z=0}=NM\nu^2\langle \Delta z_\mathrm{dom}^2\rangle /(k_BT)\big\rvert_{\delta z=0}$, where angle brackets indicate a thermal average including the tweezer bias $\delta z$ (see Supplementary Information Section IV). Thus, the susceptibility is directly related to the size of fluctuations in the system through the variance $\langle \mathrm{\Delta}z_\mathrm{dom}^2\rangle$. In a macroscopic system, one would expect $\chi$ to diverge according to $1/(1-(\mathrm{\Omega/\Omega_c})^2)$ as $\mathrm{\Omega}$ approaches $\mathrm{\Omega_c}$. Instead, because of our finite atom number, our model predicts that $\chi$ should remain finite at the critical point. 

Specifically, at the critical point, our model predicts that $\chi$ should be approximately proportional to $\sqrt{N}$ as a result of the variance $\langle \mathrm{\Delta}z_\mathrm{dom}^2\rangle$ scaling as $1/\sqrt{N}$ at leading order. This has a simple interpretation: At smaller $N$, the fluctuations in $z_\mathrm{dom}$ are larger and the atoms sample the quartic part of the effective optomechanical potential $V_\mathrm{eff}$, which provides the confinement that prevents the susceptibility from diverging. While we observe finite $\chi$ at the critical point, we are unable to resolve the $\sqrt{N}$-dependence due to the quality of our data. Nonetheless, we emphasize that this observation of finite $\chi$ is a direct measure of mesoscopic fluctuations and not just a result of non-zero temperature---a macroscopic system at non-zero temperature would still exhibit a diverging susceptibility. 
As $\mathrm{\Omega}/\mathrm{\Omega_c}$ is increased beyond 1, our model predicts that $\chi$ should continue to grow and eventually scale proportionally to $N$, before saturating at a value of $NM\nu^2\lambda^2/(16 k_B T)$. For $\mathrm{\Omega}\gg\mathrm{\Omega_c}$, the $\delta z$ required to confine $\langle z_\mathrm{dom}\rangle$ effectively to one of the minima of $V_\mathrm{eff}$ scales inversely with $N$, leading to $\chi\propto N$. Our model assumes that the atom temperature is the same for all $N$; while recoil heating may contribute to an $N$-dependent temperature and in turn affect the scaling of $\chi$ with $N$, the $1/\sqrt{N}$ scaling of $\Omega_\mathrm{c}$ observed in Fig.~\ref{fig:M2}b belies any strong $N$-dependence of the temperature. We note that 
$\chi$ continuing to increase for $\Omega>\Omega_\mathrm{c}$ is a consequence of us applying the tweezer shift $\delta z$ to break symmetry before the cavity interaction is turned on, analogous to applying a magnetic field to a spin system before lowering its temperature below the Curie point. In contrast, if the tweezer shift were applied after the cavity interaction, one might expect $\chi$ to peak at $\Omega=\Omega_\mathrm{c}$. Accordingly, we see the separation between the $N=14$ and $N=18$ data become more pronounced as we increase $\mathrm{\Omega}$ beyond $\mathrm{\Omega_c}$, consistent with the prediction of $\chi$ approaching linear scaling with $N$. These results highlight how the response of mesoscopic systems deviates from that of macroscopic systems. Furthermore, the use of tweezers to displace the atoms provides a novel method for measuring optomechanical susceptibility, complementing experiments where the external perturbation is introduced by injecting a weak field into the cavity~\cite{mottlRotonTypeModeSoftening2012,helsonDensitywaveOrderingUnitary2023}.

\section{Discussion and outlook}
We have studied self-organization, bifurcation, and symmetry breaking as well as mesoscopic dynamics and response due to cavity-mediated interactions in an atom array, opening new directions for these systems. The realization of mesoscopic self-organization with control over individual atoms has significant implications not just for our fundamental understanding of mesoscopic systems, but also for quantum simulation, optimization, and metrology. On one hand, the dynamics of the cavity field provide a plethora of information that may further our understanding of finite-size scaling~\cite{vukicsFinitesizeScalingPhotonblockade2019} and nonequilibrium physics in mesoscopic systems~\cite{diterlizziVarianceSumRule2024}, as well as pave the way for time-resolved quantum simulation~\cite{suFastSingleAtom2025,luoMomentumexchangeInteractionsBragg2024}. On the other hand, mesoscopic self-organization may also open the door to programmable engineering~\cite{periwalProgrammableInteractionsEmergent2021b} of exotic symmetry-breaking phases of matter~\cite{kroezeReplicaSymmetryBreaking2023}, for instance by illuminating groups of atoms with different phases of pump light. This system can also be used to implement a quantum numerical solver for a variety of classically hard computational problems~\cite{yeUniversalQuantumOptimization2023,torgglerQuantumNQueensSolver2019,anikeevaNumberPartitioningGrovers2021}, where the tweezer array allows for highly programmable initial conditions. Additionally, the optomechanical susceptibility near the critical point can be a resource for enhanced sensing of applied forces~\cite{iliasCriticalityEnhancedQuantumSensing2022,fernandez-lorenzoQuantumSensingClose2017,tsangQuantumTransitionedgeDetectors2013}.

\section{Acknowledgments}
We thank Nathan Song for assistance in the lab. We acknowledge support from the AFOSR (Grant No.\ FA9550-1910328 (D.M.S.-K.) and Young Investigator Prize Grant No.\ 21RT0751 (A.A.-G.), from ARO through the MURI program (Grant No.\ W911NF-20-1-0136 (D.M.S.-K.)), from DARPA (Grant No.\ W911NF2010090 (D.M.S.-K.)), from the NSF (QLCI program through grant number OMA-2016245 (D.M.S.-K.), and CAREER Award No.\ 2047380 (A.A.-G.)), from the David and Lucile Packard Foundation (A.A.-G.), and from the U.S. Department of Energy, Office of Science, National Quantum Information Science Research Centers, Quantum Systems Accelerator (D.M.S.-K.).
J.H. acknowledges support from the Department of Defense through the National Defense Science and Engineering Graduate (NDSEG) Fellowship Program. C.C.R. acknowledges support from the European Union’s Horizon Europe program under the Marie Sklodowska Curie Action LIME (Grant No. 101105916).

\textbf{Author contributions:} J.H., Y.-H.L., Z.Y., and T.X. contributed to building the experimental setup, performing the experiments, and analyzing the data. C.C.R., S.J.M., A.A.-G., D.M.S.-K., and J.H. contributed to the theoretical model. Z.Y. and D.M.S.-K. conceived the experiments. All authors contributed to the writing of the manuscript and discussed the results. All work was supervised by D.M.S.-K. and A.A.-G.

\textbf{Competing interests:} The authors declare no competing interests.

\textbf{Data and materials availability:} Data are available from the corresponding author upon reasonable request.

\section*{Methods}
\subsection*{Atom array preparation}
Our tweezer array is created by light at a wavelength of 808~nm that passes through a one-dimensional acousto-optic deflector (AOD), resembling a method described in Ref.~\cite{endresAtombyatomAssemblyDefectfree2016}. The tweezer positions are determined by radio-frequency tones, created through an arbitrary waveform generator card, that drive the AOD. The atoms are initially probabilistically loaded into a tweezer array containing 40 sites. We then perform parity projection through polarization gradient cooling (PGC) in a lin$\perp$lin configuration. The occupation of the sites is determined by fluorescence imaging. Following tweezer loading, we implement two rounds of tweezer rearrangement to bring the loaded atoms into a centered array. Unwanted atoms are dropped by diminishing the optical power in their respective tweezers in order to achieve the desired atom number between $N=10$ and $N=22$. We confirm the success of this sorting procedure with an additional fluorescence image. In the event that defects occur during the rearrangement, as noted in images taken before the pumping sequence (see the following section), we post-select our data on there being an equal number of even- and odd-indexed sites filled. We take an additional fluorescence image after the pumping sequence and retain only data where the occupancy of the pre- and post-pumping-sequence atom array is identical. This is to ensure that there is no bias in the tweezer positions towards the positive or negative antinodes when we take bifurcation data such as those shown in Fig.~1 and Fig.~2 of the main text. It is possible that this post-selection of data could leave residual bias in our analysis by selecting the most favorable shots, for instance when measuring the lifetime of the self-organized phase. However, we expect that this effect is small. To quantify, for the data shown in Figs.~1-3, the proportion of experimental runs that are kept after post-selection is 79\%. We saw single-atom loss in 12\% of the total number of runs (equivalently, 57\% of the discarded runs). Including these one-atom loss runs in our analysis leaves no visible effect on the results shown in the current figures.

The range of atom numbers $10\leq N\leq 22$ was chosen for a couple of reasons. Empirically, considering the same timescales for driving and probing the system as chosen in this work, we found that for $N<10$ the atom array was subject to significant atom loss already at pump strengths where one might expect to see self-organization; thus, $N<10$ data are not included in our analysis. The maximum atom number of $N=22$ is mostly limited by the field of view of our objective and the available laser power for the optical tweezers.

\subsection*{Pumping sequence and cavity field phase calibration}
We use heterodyne detection to measure the cavity field polarized in the $y$-direction. Our heterodyne system is schematically shown in Extended Data Fig.~\ref{fig:S1}. By demodulating the signal from the balanced photodiode at the pump-local oscillator (LO) beat frequency of 20 MHz, we extract both quadratures of the cavity field. Due to the pump and LO emerging from separate fibers, their relative phase $\phi$ drifts over time, thus necessitating calibration of the phase by referencing it to the field emitted by the atoms. To perform the phase calibration in each shot, we first demodulate the field by applying a fast Fourier transform (FFT) over a $5\,\mu \mathrm{s}$ period with a Hann window applied. The window is then stepped in $1\,\mu \mathrm{s}$ increments. The average phase, up to a $\pi$ rotation, is obtained by performing principal component analysis (PCA) on all data acquired in the shot. To calibrate the cavity field phase in each shot, we use a multi-frame pumping sequence shown in Extended Data Fig.~\ref{fig:S2} and described as follows. In frame 1, we initialize the atoms all on the cavity field antinodes with 5\,$\lambda$ spacing and pump them at a power below the critical point; we define the cavity field phase obtained here as a reference for the phase at the positive antinodes. Then, we move the tweezers into a $4.5\,\lambda$ spacing configuration while simultaneously shifting them to the cavity nodes. Next, in frame 2, we pump the atoms with variable power to detect self-organization and symmetry breaking. We take multiple frames in this configuration (typically between 5 and 20 shots) to increase our data-taking rate. Lastly, we shift the atoms back to the antinodes with 5$\,\lambda$ spacing and obtain another phase reference in the final frame of the sequence by the same method as in frame 1. We observe the phase drift between the first and last reference frames to be less than $\pi$. The sign of $c_\mathrm{proj}$ obtained in the middle frames is determined by comparing the phase extracted through PCA to the phase linearly interpolated between the two reference phases; we assume that the PCA phase is less than $\pi/2$ shifted from the interpolated phase. Example time traces of the cavity field measurement in each frame are shown in Extended Data Fig.~\ref{fig:S2}. In each frame, we linearly ramp up the pump power to the desired strength in $50\,\mathrm{\mu s}$ and then hold the power constant while simultaneously shining on repump light. Before each frame, we perform 1\,ms of PGC to set the initial temperature of the atoms. The $\sigma^-$-polarized pump beam optically pumps the atoms into the $\lvert F=2,m_F=-2 \rangle$ state with the quantization axis along $x$ during the pump ramp-up. This is done so that when $\mathrm{\Omega}<\mathrm{\Omega_c}$, the atoms stay in $\lvert F=2,m_F=-2\rangle$ and no spin dynamics interfere with the spatial self-organization process. 

\subsection*{Tweezer position calibration}
As noted in the main text, precise control of the tweezer positions with respect to the cavity and pump standing waves is crucial for our experiment. To constrain long-term drift in the position of the array along both the cavity and pump axes, we perform an in-situ measurement of both positions at end of each experimental run, and use the output of this measurement to displace the atom array onto the correct position prior to the start of the next experimental run. For the position measurement, after taking the post-pumping-sequence fluorescence image, we adjust the tweezer spacing (by varying the radio-frequency signal to the AOD) to an equal $5\,\lambda$ between neighboring tweezers. For this calibration, we use $5\,\lambda$ spacing so that the dominant mode is the center-of-mass mode, which we can easily displace using a piezo-actuated mirror in the tweezer optical path (a method described in Ref.~\cite{yanSuperradiantSubradiantCavity2023a}). This is in contrast to the $4.5\,\lambda$ spacing used in the self-organization data shots. We then expose the array to a low-intensity pump field (well below the self-organization transition) simultaneously with PGC light and record the photon emission rate from the cavity. During this recording, the entire array is translated continuously first along the cavity axis and then along the pump axis using the piezo-actuated mirror. The recorded data reveal the sinusoidal variation of the cavity mode and the pump standing wave, respectively. An automated fit applied to these data determines the displacement that must be applied to the atoms to position them properly within the cavity and pump field.

\subsection*{Experimental interpolation of the critical point: data and error analysis}
As noted in the main text, the critical point is determined by interpolating the value of $\mathrm{\Omega}$ that makes $B=0$, where $k_BT\times B$ is the coefficient on the quadratic term of the mechanical potential. At each parameter setting, we fit the obtained distributions of $c_\mathrm{proj}$ to the Boltzmann distribution (Equation~(\ref{z_dist})) to extract $B$. Examples of such distributions and their Boltzmann fits are shown in Extended Data Fig.~\ref{fig:S4}a. To estimate the error in this fit of $B$, we implement bootstrapping: we randomly draw samples with replacement from the $c_\mathrm{proj}$ distribution at each $\Omega$ and do the fit again. This is iterated 10 times. The average value and standard deviation of $B$ is then calculated from the bootstrap samples.  The sign of $B$ determines whether the distribution is unimodal ($B>0$) or bimodal ($B<0$), and thus the critical point occurs when $B=0$. As shown in Extended Data Fig.~\ref{fig:S4}b, the fitted values of $B$ go from being positive to negative as $\mathrm{\Omega}$ is increased. 

Since we lack a complete theoretical model of how $B$ should vary with $\mathrm{\Omega}$, we simply linearly interpolate the value of $\mathrm{\Omega}$ that makes $B=0$ and use this as $\mathrm{\Omega_c}$. To do so, we consider the two sampled points directly on either side of $B=0$; we call them by the coordinates ($\mathrm{\Omega}_1,B_1$) and ($\mathrm{\Omega}_2,B_2$), where $B_1>0$ and $B_2<0$. Then, our experimental critical point $\Omega_\mathrm{c}^\mathrm{(exp)}$ is given by the $x$-intercept of the line that passes through these two points: 
\begin{equation}
    \label{Omega_c_exp}
    \mathrm{\Omega_c}^{(\mathrm{exp})}=\mathrm{\Omega_1}-\frac{B_1(\mathrm{\Omega_1}-\mathrm{\Omega_2})}{B_1-B_2}.
\end{equation}
The standard deviation $\sigma_\mathrm{\Omega_c}$ on the critical point is obtained through error propagation: 
\begin{eqnarray}
    \sigma_\mathrm{\Omega_c}^2=\Big\lvert\frac{\partial\mathrm{\Omega_c}^{(\mathrm{exp})}}{\partial B_1}\Big\rvert^2\sigma_{B_1}^2+\Big\lvert\frac{\partial\mathrm{\Omega_c}^{(\mathrm{exp})}}{\partial B_2}\Big\rvert^2\sigma_{B_2}^2\\\nonumber+\Big\lvert\frac{\partial\mathrm{\Omega_c}^{(\mathrm{exp})}}{\partial \mathrm{\Omega}_1}\Big\rvert^2\sigma_{\mathrm{\Omega}_1}^2+\Big\lvert\frac{\partial\mathrm{\Omega_c}^{(\mathrm{exp})}}{\partial \mathrm{\Omega}_2}\Big\rvert^2\sigma_{\mathrm{\Omega}_2}^2.
\end{eqnarray}
The standard deviations $\sigma_{B_1}$ and $\sigma_{B_2}$ are obtained from bootstrapping, while $\sigma_{\mathrm{\Omega}_1}$ and $\sigma_{\mathrm{\Omega}_2}$ are given by an estimate of 10\% drift of the pump power over the course of the experimental runs. This estimation is based off measurements of the pump power at the start and end of data-taking. We note that the linear interpolation is an approximate method for obtaining the critical point, and so the error propagation applied to obtain $\sigma_{\Omega_\mathrm{c}}$ may not accurately reflect the true experimental uncertainty on $\Omega_\mathrm{c}$.

Our extraction of $\Omega_\mathrm{c}^\mathrm{(exp)}$ is also subject to systematic errors due to Gaussian noise in our detection and uncertainty in our measurement of the pump Rabi frequency. Gaussian noise, resulting from photon shot noise and electronic detector noise, causes the $c_\mathrm{proj}$ distributions to smear out and thus leads us to systematically overestimate $\Omega_\mathrm{c}^\mathrm{(exp)}$ when we use the method described above. We theoretically calculate how much the noise affects $\Omega_\mathrm{c}^\mathrm{(exp)}$ by convolving a Gaussian distribution, whose width is determined by the fluctuations observed when no atoms are in the cavity, with the Boltzmann distribution. We then perform the same $B$-fits and linear interpolation on the convolved distributions to estimate the percentage by which $\Omega_\mathrm{c}^\mathrm{(exp)}$ changes. All $\Omega_\mathrm{c}$ data shown in the main text and Extended Data Fig.~\ref{fig:S3} have been scaled to account for this systematic error.

The other systematic error we have identified is error in our measurement of $\Omega$ due to uncertainty in our measurement of the pump beam size at the atoms ($173\pm5\,\mu$m), in the positioning of the pump beam relative to the atom array, and in our measurement of the pump power (estimated to be $\pm 6$\%). Since the same beam size and power was applied to all data, these will cause all the $\Omega_\mathrm{c}$ data to be shifted together up to approximately $\pm 6$\%. We also note that the average $\Omega$ seen by the atom array depends on the atom number, since a larger array will see more of the Gaussian profile of the pump beam. To account for this, we average the pump intensity profile over the span of the atom array for each $N$, assuming the beam is exactly centered on the array, and factor this into our reported $\Omega$; this systematically shifts $\Omega$ downwards. It is also possible that the beam is not exactly centered on the atoms; we estimate the uncertainty to be about $\pm30\,\mu$m in both the $y$ and $z$ directions. This off-centering would reduce $\Omega$ by up to an additional 6\%. As a result, we expect that $\Omega_\mathrm{c}^\mathrm{(exp)}$ could be systematically shifted upwards by up to 6\% or shifted downwards by up to 8.5\%, where we have added the downward errors in quadrature.

There are additional systematic error sources whose effects on the critical point have not been quantified. One such source is the tweezer trap frequency being lower at the outer edges of the array due to aberrations in the tweezer optical setup. This should push the critical point lower for larger $N$. Similarly, optical aberrations make the tweezer spacing uneven towards the edges of the array; we estimate the deviation to be up to 100\,nm along both the cavity and pump axes at the edges of the 22-atom array. We expect this deviation to push the critical point higher for larger atom numbers.

\renewcommand{\figurename}{Extended Data Fig.}

\setcounter{figure}{0}

\begin{figure*}
    \centering
    \includegraphics[width=0.8\textwidth]{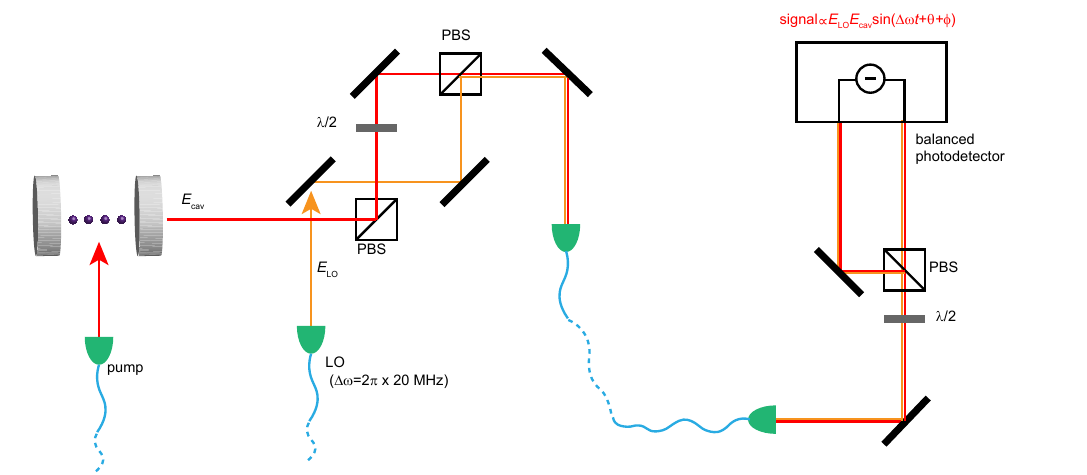}
    \caption{\textbf{Schematic of the heterodyne detection system.} The pump and LO beams originate from separate fibers and are combined at a polarizing beamsplitter (PBS) with orthogonal polarizations. They are then both sent into a single mode fiber, which ensures that they are mode-matched at the fiber output. Upon emerging from the fiber, they pass through a half-waveplate (labeled as $\lambda/2$) at 45$^\circ$ and are split at another PBS. Each output port of the last PBS is directed to a sensor on a balanced photodetector, which subtracts the signals on the two sensors. The output voltage of the detector is proportional to $E_\mathrm{LO}E_\mathrm{cav}\sin(\mathrm{\Delta}\omega t+\vartheta+\phi)$ where $E_\mathrm{LO}$ is the LO electric field amplitude, $E_\mathrm{cav}$ is the cavity electric field amplitude, $\mathrm{\Delta}\omega=2\pi\times 20\,\mathrm{MHz}$ is the frequency difference between the pump and LO, $\phi$ is the phase of the LO relative to the pump, and $\theta$ is the phase of the cavity field relative to the pump.}
    \label{fig:S1}
\end{figure*}

\begin{figure*}
    \centering
    \includegraphics[width=0.8\textwidth]{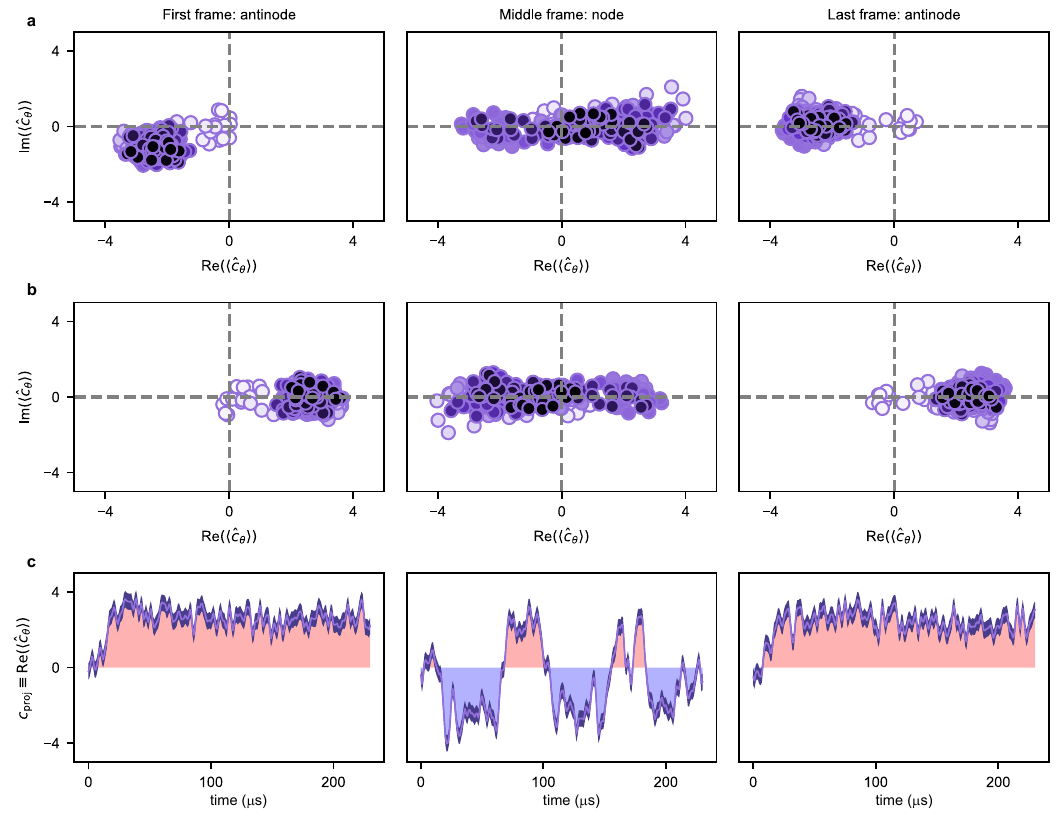}
    \caption{\textbf{Example time traces of the detected field in each of the three pumping frames.} \textbf{a,} Scatter plots of the complex cavity field before phase correction. The real ($\mathrm{Re}(\langle\hat{c}_\vartheta\rangle)$) and imaginary ($\mathrm{Im}(\langle\hat{c}_\vartheta\rangle)$) quadratures of the detected field with respect to the average phase of the cavity emission, $\vartheta$, are plotted. Lighter points indicate earlier times and darker points indicate later times. From these data, the angle $\phi$ of the major axis in each frame is determined by PCA. \textbf{b,} Detected field after rotation by $\phi$. Because PCA only determines the angle of the major axis up to a $\pi$ rotation, in the antinode frames, we define the average projection of the field onto the axis defined by $\vartheta$ to be positive. The sign of the node frame is determined by comparing $\vartheta_\mathrm{node}$ to the angle linearly interpolated from $\vartheta_\text{first frame}$ and $\vartheta_\text{last frame}$ and assuming that the two angles are within $\pi/2$ of each other. \textbf{c,} Time traces of $c_\mathrm{proj}$. Values of $c_\mathrm{proj}$ are obtained by projecting the phase-corrected measurements of $\hat{c}$ (shown in (\textbf{b})) onto the $\vartheta$-axis. The shaded dark purple regions indicate the shot noise level. Data shown in figure were taken with $N=18$, $\mathrm{\Delta_{pa}}=-2\pi\times 80\,\mathrm{MHz}$, $\mathrm{\Delta_{pc}}=-2\pi\times 2.15\,\mathrm{MHz}$, and $\mathrm{\Omega}=2\pi\times 63.9\,\mathrm{MHz}$.}
    \label{fig:S2}
\end{figure*}

\begin{figure*}
    \centering
    \includegraphics[width=0.8\textwidth]{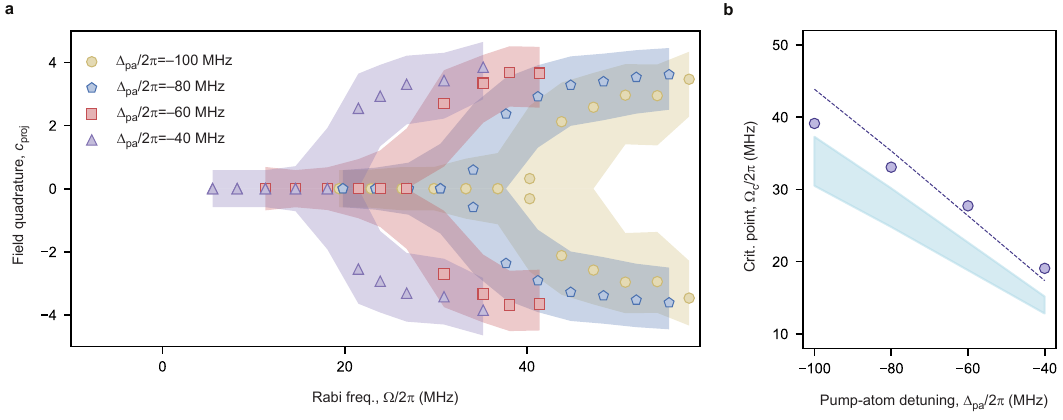}
    \caption{\textbf{Dependence of the critical point on $\mathrm{\Delta_{pa}}$.} \textbf{a,} Data showing the bifurcation for 20 atoms at four values of $\mathrm{\Delta_{pa}}$. Markers represent the maxima of the fitted Boltzmann probability distribution of $c_\mathrm{proj}$. Shaded areas show the width of the distribution at half-maximum. Different pump-cavity detunings were chosen for each setting of $\mathrm{\Delta_{pa}}$ to compensate for the different dispersive shifts on the cavity resonance frequency for 20 atoms placed at the nodes. The values used are $\mathrm{\Delta_{pc}}=-2\pi\times \{1.79, 1.9, 2.02, 2.26\}\,\mathrm{MHz}$ corresponding to $\mathrm{\Delta_{pa}}=-2\pi\times \{100, 80, 60, 40\}\,\mathrm{MHz}$. These correspond approximately to $\mathrm{\Delta_{pc}}(T)\simeq-2\pi\times 1.6\,\mathrm{MHz}$ when accounting for the shift that thermal atoms put on the cavity resonance frequency. \textbf{b,} Extracted $\mathrm{\Omega_c}$ (circles) exhibit an approximately linear dependence on $\mathrm{\Delta_{pa}}$.  The dashed line shows a fit to Equation~(\ref{Omega_c}), which gives a fitted temperature of $T=66\pm8\,\mu$K. The shaded region shows the prediction of Equation~(\ref{Omega_c}) for the independent temperature measurement. Error bars are smaller than the markers.}
    \label{fig:S3}
\end{figure*}

\begin{figure*}
    \centering
    \includegraphics[width=0.8\textwidth]{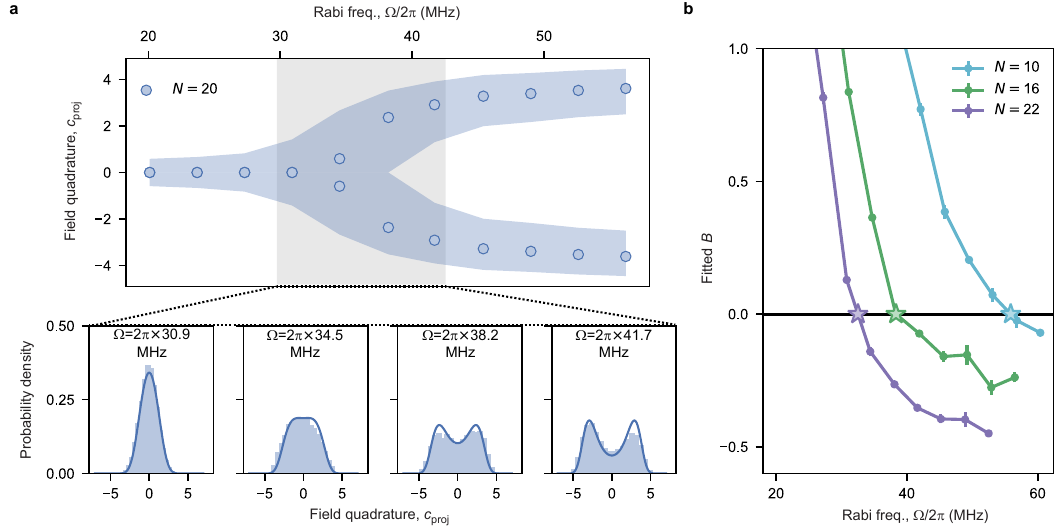}
    \caption{\textbf{Interpolation of $\mathrm{\Omega_c}$ from Boltzmann fits.} \textbf{a,} Data showing the bifurcation for 20 atoms at  $\mathrm{\Delta_{pa}}=2\pi\times 80\,\mathrm{MHz}$. Markers represent the maxima of the fitted Boltzmann probability distribution of $c_\mathrm{proj}$. Blue shaded areas show the width of the distribution at half-maximum. Bottom insets show histograms (light blue) and Boltzmann fits (dark blue) for values of $\mathrm{\Omega}$ in the gray shaded region.
    \textbf{b,} By fitting experimentally obtained distributions, such as those shown in (\textbf{a}), to the Boltzmann distribution, we extract the parameter $B$ at different values of $\mathrm{\Omega}$ for various $N$. The $N$ shown here correspond to the data shown in Fig.\,2a of the main text. The value of $\mathrm{\Omega_c}$ (stars) is linearly interpolated from the two points on directly on either side of the $B=0$ line. Error bars on the fitted $B$ values are smaller than the markers. Lines are guides to the eye.}
    \label{fig:S4}
\end{figure*}

\clearpage
\bibliography{library}

\begin{thebibliography}{61}%
\makeatletter
\providecommand \@ifxundefined [1]{%
 \@ifx{#1\undefined}
}%
\providecommand \@ifnum [1]{%
 \ifnum #1\expandafter \@firstoftwo
 \else \expandafter \@secondoftwo
 \fi
}%
\providecommand \@ifx [1]{%
 \ifx #1\expandafter \@firstoftwo
 \else \expandafter \@secondoftwo
 \fi
}%
\providecommand \natexlab [1]{#1}%
\providecommand \enquote  [1]{``#1''}%
\providecommand \bibnamefont  [1]{#1}%
\providecommand \bibfnamefont [1]{#1}%
\providecommand \citenamefont [1]{#1}%
\providecommand \href@noop [0]{\@secondoftwo}%
\providecommand \href [0]{\begingroup \@sanitize@url \@href}%
\providecommand \@href[1]{\@@startlink{#1}\@@href}%
\providecommand \@@href[1]{\endgroup#1\@@endlink}%
\providecommand \@sanitize@url [0]{\catcode `\\12\catcode `\$12\catcode `\&12\catcode `\#12\catcode `\^12\catcode `\_12\catcode `\%12\relax}%
\providecommand \@@startlink[1]{}%
\providecommand \@@endlink[0]{}%
\providecommand \url  [0]{\begingroup\@sanitize@url \@url }%
\providecommand \@url [1]{\endgroup\@href {#1}{\urlprefix }}%
\providecommand \urlprefix  [0]{URL }%
\providecommand \Eprint [0]{\href }%
\providecommand \doibase [0]{https://doi.org/}%
\providecommand \selectlanguage [0]{\@gobble}%
\providecommand \bibinfo  [0]{\@secondoftwo}%
\providecommand \bibfield  [0]{\@secondoftwo}%
\providecommand \translation [1]{[#1]}%
\providecommand \BibitemOpen [0]{}%
\providecommand \bibitemStop [0]{}%
\providecommand \bibitemNoStop [0]{.\EOS\space}%
\providecommand \EOS [0]{\spacefactor3000\relax}%
\providecommand \BibitemShut  [1]{\csname bibitem#1\endcsname}%
\let\auto@bib@innerbib\@empty
\bibitem [{\citenamefont {Anderson}(1972)}]{andersonMoreDifferent1972}%
  \BibitemOpen
  \bibfield  {author} {\bibinfo {author} {\bibfnamefont {P.}~\bibnamefont {Anderson}},\ }\bibfield  {title} {\bibinfo {title} {More {{Is Different}}},\ }\href@noop {} {\bibfield  {journal} {\bibinfo  {journal} {Science}\ }\textbf {\bibinfo {volume} {177}},\ \bibinfo {pages} {393} (\bibinfo {year} {1972})}\BibitemShut {NoStop}%
\bibitem [{\citenamefont {Jou}\ and\ \citenamefont {Restuccia}(2011)}]{jouMesoscopicTransportEquations2011}%
  \BibitemOpen
  \bibfield  {author} {\bibinfo {author} {\bibfnamefont {D.}~\bibnamefont {Jou}}\ and\ \bibinfo {author} {\bibfnamefont {L.}~\bibnamefont {Restuccia}},\ }\bibfield  {title} {\bibinfo {title} {Mesoscopic transport equations and contemporary thermodynamics: An introduction},\ }\href {https://doi.org/10.1080/00107514.2011.595596} {\bibfield  {journal} {\bibinfo  {journal} {Contemporary Physics}\ }\textbf {\bibinfo {volume} {52}},\ \bibinfo {pages} {465} (\bibinfo {year} {2011})}\BibitemShut {NoStop}%
\bibitem [{\citenamefont {Iachello}\ and\ \citenamefont {Zamfir}(2004)}]{iachelloQuantumPhaseTransitions2004}%
  \BibitemOpen
  \bibfield  {author} {\bibinfo {author} {\bibfnamefont {F.}~\bibnamefont {Iachello}}\ and\ \bibinfo {author} {\bibfnamefont {N.~V.}\ \bibnamefont {Zamfir}},\ }\bibfield  {title} {\bibinfo {title} {Quantum {{Phase Transitions}} in {{Mesoscopic Systems}}},\ }\href {https://doi.org/10.1103/PhysRevLett.92.212501} {\bibfield  {journal} {\bibinfo  {journal} {Physical Review Letters}\ }\textbf {\bibinfo {volume} {92}},\ \bibinfo {pages} {212501} (\bibinfo {year} {2004})}\BibitemShut {NoStop}%
\bibitem [{\citenamefont {Vidal}\ and\ \citenamefont {Dusuel}(2006)}]{vidalFinitesizeScalingExponents2006}%
  \BibitemOpen
  \bibfield  {author} {\bibinfo {author} {\bibfnamefont {J.}~\bibnamefont {Vidal}}\ and\ \bibinfo {author} {\bibfnamefont {S.}~\bibnamefont {Dusuel}},\ }\bibfield  {title} {\bibinfo {title} {Finite-size scaling exponents in the {{Dicke}} model},\ }\href {https://doi.org/10.1209/epl/i2006-10041-9} {\bibfield  {journal} {\bibinfo  {journal} {Europhysics Letters (EPL)}\ }\textbf {\bibinfo {volume} {74}},\ \bibinfo {pages} {817} (\bibinfo {year} {2006})}\BibitemShut {NoStop}%
\bibitem [{\citenamefont {Stitely}\ \emph {et~al.}(2020)\citenamefont {Stitely}, \citenamefont {Masson}, \citenamefont {Giraldo}, \citenamefont {Krauskopf},\ and\ \citenamefont {Parkins}}]{stitelySuperradiantSwitchingQuantum2020}%
  \BibitemOpen
  \bibfield  {author} {\bibinfo {author} {\bibfnamefont {K.~C.}\ \bibnamefont {Stitely}}, \bibinfo {author} {\bibfnamefont {S.~J.}\ \bibnamefont {Masson}}, \bibinfo {author} {\bibfnamefont {A.}~\bibnamefont {Giraldo}}, \bibinfo {author} {\bibfnamefont {B.}~\bibnamefont {Krauskopf}},\ and\ \bibinfo {author} {\bibfnamefont {S.}~\bibnamefont {Parkins}},\ }\bibfield  {title} {\bibinfo {title} {Superradiant switching, quantum hysteresis, and oscillations in a generalized {{Dicke}} model},\ }\href {https://doi.org/10.1103/PhysRevA.102.063702} {\bibfield  {journal} {\bibinfo  {journal} {Physical Review A}\ }\textbf {\bibinfo {volume} {102}},\ \bibinfo {pages} {063702} (\bibinfo {year} {2020})}\BibitemShut {NoStop}%
\bibitem [{\citenamefont {Karapetrov}\ \emph {et~al.}(2005)\citenamefont {Karapetrov}, \citenamefont {Fedor}, \citenamefont {Iavarone}, \citenamefont {Rosenmann},\ and\ \citenamefont {Kwok}}]{karapetrovDirectObservationGeometrical2005}%
  \BibitemOpen
  \bibfield  {author} {\bibinfo {author} {\bibfnamefont {G.}~\bibnamefont {Karapetrov}}, \bibinfo {author} {\bibfnamefont {J.}~\bibnamefont {Fedor}}, \bibinfo {author} {\bibfnamefont {M.}~\bibnamefont {Iavarone}}, \bibinfo {author} {\bibfnamefont {D.}~\bibnamefont {Rosenmann}},\ and\ \bibinfo {author} {\bibfnamefont {W.~K.}\ \bibnamefont {Kwok}},\ }\bibfield  {title} {\bibinfo {title} {Direct {{Observation}} of {{Geometrical Phase Transitions}} in {{Mesoscopic Superconductors}} by {{Scanning Tunneling Microscopy}}},\ }\href {https://doi.org/10.1103/PhysRevLett.95.167002} {\bibfield  {journal} {\bibinfo  {journal} {Physical Review Letters}\ }\textbf {\bibinfo {volume} {95}},\ \bibinfo {pages} {167002} (\bibinfo {year} {2005})}\BibitemShut {NoStop}%
\bibitem [{\citenamefont {Bird}\ \emph {et~al.}(1997)\citenamefont {Bird}, \citenamefont {Ishibashi}, \citenamefont {Aoyagi}, \citenamefont {Sugano}, \citenamefont {Akis}, \citenamefont {Ferry}, \citenamefont {Pivin}, \citenamefont {Connolly}, \citenamefont {Taylor}, \citenamefont {Newbury}, \citenamefont {Olatona}, \citenamefont {Micolich}, \citenamefont {Wirtz}, \citenamefont {Ochiai},\ and\ \citenamefont {Okubo}}]{birdQuantumTransportOpen1997}%
  \BibitemOpen
  \bibfield  {author} {\bibinfo {author} {\bibfnamefont {J.}~\bibnamefont {Bird}}, \bibinfo {author} {\bibfnamefont {K.}~\bibnamefont {Ishibashi}}, \bibinfo {author} {\bibfnamefont {Y.}~\bibnamefont {Aoyagi}}, \bibinfo {author} {\bibfnamefont {T.}~\bibnamefont {Sugano}}, \bibinfo {author} {\bibfnamefont {R.}~\bibnamefont {Akis}}, \bibinfo {author} {\bibfnamefont {D.}~\bibnamefont {Ferry}}, \bibinfo {author} {\bibfnamefont {D.}~\bibnamefont {Pivin}}, \bibinfo {author} {\bibfnamefont {K.}~\bibnamefont {Connolly}}, \bibinfo {author} {\bibfnamefont {R.}~\bibnamefont {Taylor}}, \bibinfo {author} {\bibfnamefont {R.}~\bibnamefont {Newbury}}, \bibinfo {author} {\bibfnamefont {D.}~\bibnamefont {Olatona}}, \bibinfo {author} {\bibfnamefont {A.}~\bibnamefont {Micolich}}, \bibinfo {author} {\bibfnamefont {R.}~\bibnamefont {Wirtz}}, \bibinfo {author} {\bibfnamefont {Y.}~\bibnamefont {Ochiai}},\ and\ \bibinfo {author} {\bibfnamefont {Y.}~\bibnamefont {Okubo}},\ }\bibfield  {title} {\bibinfo {title} {Quantum transport in
  open mesoscopic cavities},\ }\href {https://doi.org/10.1016/S0960-0779(97)00021-0} {\bibfield  {journal} {\bibinfo  {journal} {Chaos, Solitons \& Fractals}\ }\textbf {\bibinfo {volume} {8}},\ \bibinfo {pages} {1299} (\bibinfo {year} {1997})}\BibitemShut {NoStop}%
\bibitem [{\citenamefont {Knoll}\ \emph {et~al.}(2004)\citenamefont {Knoll}, \citenamefont {Lyakhova}, \citenamefont {Horvat}, \citenamefont {Krausch}, \citenamefont {Sevink}, \citenamefont {Zvelindovsky},\ and\ \citenamefont {Magerle}}]{knollDirectImagingMesoscale2004}%
  \BibitemOpen
  \bibfield  {author} {\bibinfo {author} {\bibfnamefont {A.}~\bibnamefont {Knoll}}, \bibinfo {author} {\bibfnamefont {K.~S.}\ \bibnamefont {Lyakhova}}, \bibinfo {author} {\bibfnamefont {A.}~\bibnamefont {Horvat}}, \bibinfo {author} {\bibfnamefont {G.}~\bibnamefont {Krausch}}, \bibinfo {author} {\bibfnamefont {G.~J.~A.}\ \bibnamefont {Sevink}}, \bibinfo {author} {\bibfnamefont {A.~V.}\ \bibnamefont {Zvelindovsky}},\ and\ \bibinfo {author} {\bibfnamefont {R.}~\bibnamefont {Magerle}},\ }\bibfield  {title} {\bibinfo {title} {Direct imaging and mesoscale modelling of phase transitions in a nanostructured fluid},\ }\href {https://doi.org/10.1038/nmat1258} {\bibfield  {journal} {\bibinfo  {journal} {Nature Materials}\ }\textbf {\bibinfo {volume} {3}},\ \bibinfo {pages} {886} (\bibinfo {year} {2004})}\BibitemShut {NoStop}%
\bibitem [{\citenamefont {Matheny}\ \emph {et~al.}(2019)\citenamefont {Matheny}, \citenamefont {Emenheiser}, \citenamefont {Fon}, \citenamefont {Chapman}, \citenamefont {Salova}, \citenamefont {Rohden}, \citenamefont {Li}, \citenamefont {Hudoba De~Badyn}, \citenamefont {P{\'o}sfai}, \citenamefont {{Duenas-Osorio}}, \citenamefont {Mesbahi}, \citenamefont {Crutchfield}, \citenamefont {Cross}, \citenamefont {D'Souza},\ and\ \citenamefont {Roukes}}]{mathenyExoticStatesSimple2019}%
  \BibitemOpen
  \bibfield  {author} {\bibinfo {author} {\bibfnamefont {M.~H.}\ \bibnamefont {Matheny}}, \bibinfo {author} {\bibfnamefont {J.}~\bibnamefont {Emenheiser}}, \bibinfo {author} {\bibfnamefont {W.}~\bibnamefont {Fon}}, \bibinfo {author} {\bibfnamefont {A.}~\bibnamefont {Chapman}}, \bibinfo {author} {\bibfnamefont {A.}~\bibnamefont {Salova}}, \bibinfo {author} {\bibfnamefont {M.}~\bibnamefont {Rohden}}, \bibinfo {author} {\bibfnamefont {J.}~\bibnamefont {Li}}, \bibinfo {author} {\bibfnamefont {M.}~\bibnamefont {Hudoba De~Badyn}}, \bibinfo {author} {\bibfnamefont {M.}~\bibnamefont {P{\'o}sfai}}, \bibinfo {author} {\bibfnamefont {L.}~\bibnamefont {{Duenas-Osorio}}}, \bibinfo {author} {\bibfnamefont {M.}~\bibnamefont {Mesbahi}}, \bibinfo {author} {\bibfnamefont {J.~P.}\ \bibnamefont {Crutchfield}}, \bibinfo {author} {\bibfnamefont {M.~C.}\ \bibnamefont {Cross}}, \bibinfo {author} {\bibfnamefont {R.~M.}\ \bibnamefont {D'Souza}},\ and\ \bibinfo {author} {\bibfnamefont {M.~L.}\ \bibnamefont {Roukes}},\ }\bibfield
  {title} {\bibinfo {title} {Exotic states in a simple network of nanoelectromechanical oscillators},\ }\href {https://doi.org/10.1126/science.aav7932} {\bibfield  {journal} {\bibinfo  {journal} {Science}\ }\textbf {\bibinfo {volume} {363}},\ \bibinfo {pages} {eaav7932} (\bibinfo {year} {2019})}\BibitemShut {NoStop}%
\bibitem [{\citenamefont {Wollack}\ \emph {et~al.}(2022)\citenamefont {Wollack}, \citenamefont {Cleland}, \citenamefont {Gruenke}, \citenamefont {Wang}, \citenamefont {{Arrangoiz-Arriola}},\ and\ \citenamefont {{Safavi-Naeini}}}]{wollackQuantumStatePreparation2022}%
  \BibitemOpen
  \bibfield  {author} {\bibinfo {author} {\bibfnamefont {E.~A.}\ \bibnamefont {Wollack}}, \bibinfo {author} {\bibfnamefont {A.~Y.}\ \bibnamefont {Cleland}}, \bibinfo {author} {\bibfnamefont {R.~G.}\ \bibnamefont {Gruenke}}, \bibinfo {author} {\bibfnamefont {Z.}~\bibnamefont {Wang}}, \bibinfo {author} {\bibfnamefont {P.}~\bibnamefont {{Arrangoiz-Arriola}}},\ and\ \bibinfo {author} {\bibfnamefont {A.~H.}\ \bibnamefont {{Safavi-Naeini}}},\ }\bibfield  {title} {\bibinfo {title} {Quantum state preparation and tomography of entangled mechanical resonators},\ }\href {https://doi.org/10.1038/s41586-022-04500-y} {\bibfield  {journal} {\bibinfo  {journal} {Nature}\ }\textbf {\bibinfo {volume} {604}},\ \bibinfo {pages} {463} (\bibinfo {year} {2022})}\BibitemShut {NoStop}%
\bibitem [{\citenamefont {Mirhosseini}\ \emph {et~al.}(2020)\citenamefont {Mirhosseini}, \citenamefont {Sipahigil}, \citenamefont {Kalaee},\ and\ \citenamefont {Painter}}]{mirhosseiniSuperconductingQubitOptical2020}%
  \BibitemOpen
  \bibfield  {author} {\bibinfo {author} {\bibfnamefont {M.}~\bibnamefont {Mirhosseini}}, \bibinfo {author} {\bibfnamefont {A.}~\bibnamefont {Sipahigil}}, \bibinfo {author} {\bibfnamefont {M.}~\bibnamefont {Kalaee}},\ and\ \bibinfo {author} {\bibfnamefont {O.}~\bibnamefont {Painter}},\ }\bibfield  {title} {\bibinfo {title} {Superconducting qubit to optical photon transduction},\ }\href {https://doi.org/10.1038/s41586-020-3038-6} {\bibfield  {journal} {\bibinfo  {journal} {Nature}\ }\textbf {\bibinfo {volume} {588}},\ \bibinfo {pages} {599} (\bibinfo {year} {2020})}\BibitemShut {NoStop}%
\bibitem [{\citenamefont {Ren}\ \emph {et~al.}(2022)\citenamefont {Ren}, \citenamefont {Shah}, \citenamefont {Pfeifer}, \citenamefont {Brendel}, \citenamefont {Peano}, \citenamefont {Marquardt},\ and\ \citenamefont {Painter}}]{renTopologicalPhononTransport2022}%
  \BibitemOpen
  \bibfield  {author} {\bibinfo {author} {\bibfnamefont {H.}~\bibnamefont {Ren}}, \bibinfo {author} {\bibfnamefont {T.}~\bibnamefont {Shah}}, \bibinfo {author} {\bibfnamefont {H.}~\bibnamefont {Pfeifer}}, \bibinfo {author} {\bibfnamefont {C.}~\bibnamefont {Brendel}}, \bibinfo {author} {\bibfnamefont {V.}~\bibnamefont {Peano}}, \bibinfo {author} {\bibfnamefont {F.}~\bibnamefont {Marquardt}},\ and\ \bibinfo {author} {\bibfnamefont {O.}~\bibnamefont {Painter}},\ }\bibfield  {title} {\bibinfo {title} {Topological phonon transport in an optomechanical system},\ }\href {https://doi.org/10.1038/s41467-022-30941-0} {\bibfield  {journal} {\bibinfo  {journal} {Nature Communications}\ }\textbf {\bibinfo {volume} {13}},\ \bibinfo {pages} {3476} (\bibinfo {year} {2022})}\BibitemShut {NoStop}%
\bibitem [{\citenamefont {Andrade}\ \emph {et~al.}(2006)\citenamefont {Andrade}, \citenamefont {Laraoui}, \citenamefont {Vomir}, \citenamefont {Muller}, \citenamefont {Stoquert}, \citenamefont {Estourn{\`e}s}, \citenamefont {Beaurepaire},\ and\ \citenamefont {Bigot}}]{andradeDampedPrecessionMagnetization2006}%
  \BibitemOpen
  \bibfield  {author} {\bibinfo {author} {\bibfnamefont {L.~H.~F.}\ \bibnamefont {Andrade}}, \bibinfo {author} {\bibfnamefont {A.}~\bibnamefont {Laraoui}}, \bibinfo {author} {\bibfnamefont {M.}~\bibnamefont {Vomir}}, \bibinfo {author} {\bibfnamefont {D.}~\bibnamefont {Muller}}, \bibinfo {author} {\bibfnamefont {J.-P.}\ \bibnamefont {Stoquert}}, \bibinfo {author} {\bibfnamefont {C.}~\bibnamefont {Estourn{\`e}s}}, \bibinfo {author} {\bibfnamefont {E.}~\bibnamefont {Beaurepaire}},\ and\ \bibinfo {author} {\bibfnamefont {J.-Y.}\ \bibnamefont {Bigot}},\ }\bibfield  {title} {\bibinfo {title} {Damped {{Precession}} of the {{Magnetization Vector}} of {{Superparamagnetic Nanoparticles Excited}} by {{Femtosecond Optical Pulses}}},\ }\href {https://doi.org/10.1103/PhysRevLett.97.127401} {\bibfield  {journal} {\bibinfo  {journal} {Physical Review Letters}\ }\textbf {\bibinfo {volume} {97}},\ \bibinfo {pages} {127401} (\bibinfo {year} {2006})}\BibitemShut {NoStop}%
\bibitem [{\citenamefont {Thirion}\ \emph {et~al.}(2003)\citenamefont {Thirion}, \citenamefont {Wernsdorfer},\ and\ \citenamefont {Mailly}}]{thirionSwitchingMagnetizationNonlinear2003}%
  \BibitemOpen
  \bibfield  {author} {\bibinfo {author} {\bibfnamefont {C.}~\bibnamefont {Thirion}}, \bibinfo {author} {\bibfnamefont {W.}~\bibnamefont {Wernsdorfer}},\ and\ \bibinfo {author} {\bibfnamefont {D.}~\bibnamefont {Mailly}},\ }\bibfield  {title} {\bibinfo {title} {Switching of magnetization by nonlinear resonance studied in single nanoparticles},\ }\href {https://doi.org/10.1038/nmat946} {\bibfield  {journal} {\bibinfo  {journal} {Nature Materials}\ }\textbf {\bibinfo {volume} {2}},\ \bibinfo {pages} {524} (\bibinfo {year} {2003})}\BibitemShut {NoStop}%
\bibitem [{\citenamefont {Bayha}\ \emph {et~al.}(2020)\citenamefont {Bayha}, \citenamefont {Holten}, \citenamefont {Klemt}, \citenamefont {Subramanian}, \citenamefont {Bjerlin}, \citenamefont {Reimann}, \citenamefont {Bruun}, \citenamefont {Preiss},\ and\ \citenamefont {Jochim}}]{bayhaObservingEmergenceQuantum2020}%
  \BibitemOpen
  \bibfield  {author} {\bibinfo {author} {\bibfnamefont {L.}~\bibnamefont {Bayha}}, \bibinfo {author} {\bibfnamefont {M.}~\bibnamefont {Holten}}, \bibinfo {author} {\bibfnamefont {R.}~\bibnamefont {Klemt}}, \bibinfo {author} {\bibfnamefont {K.}~\bibnamefont {Subramanian}}, \bibinfo {author} {\bibfnamefont {J.}~\bibnamefont {Bjerlin}}, \bibinfo {author} {\bibfnamefont {S.~M.}\ \bibnamefont {Reimann}}, \bibinfo {author} {\bibfnamefont {G.~M.}\ \bibnamefont {Bruun}}, \bibinfo {author} {\bibfnamefont {P.~M.}\ \bibnamefont {Preiss}},\ and\ \bibinfo {author} {\bibfnamefont {S.}~\bibnamefont {Jochim}},\ }\bibfield  {title} {\bibinfo {title} {Observing the emergence of a quantum phase transition shell by shell},\ }\href {https://doi.org/10.1038/s41586-020-2936-y} {\bibfield  {journal} {\bibinfo  {journal} {Nature}\ }\textbf {\bibinfo {volume} {587}},\ \bibinfo {pages} {583} (\bibinfo {year} {2020})}\BibitemShut {NoStop}%
\bibitem [{\citenamefont {Brantut}\ \emph {et~al.}(2012)\citenamefont {Brantut}, \citenamefont {Meineke}, \citenamefont {Stadler}, \citenamefont {Krinner},\ and\ \citenamefont {Esslinger}}]{brantutConductionUltracoldFermions2012}%
  \BibitemOpen
  \bibfield  {author} {\bibinfo {author} {\bibfnamefont {J.-P.}\ \bibnamefont {Brantut}}, \bibinfo {author} {\bibfnamefont {J.}~\bibnamefont {Meineke}}, \bibinfo {author} {\bibfnamefont {D.}~\bibnamefont {Stadler}}, \bibinfo {author} {\bibfnamefont {S.}~\bibnamefont {Krinner}},\ and\ \bibinfo {author} {\bibfnamefont {T.}~\bibnamefont {Esslinger}},\ }\bibfield  {title} {\bibinfo {title} {Conduction of {{Ultracold Fermions Through}} a {{Mesoscopic Channel}}},\ }\href {https://doi.org/10.1126/science.1223175} {\bibfield  {journal} {\bibinfo  {journal} {Science}\ }\textbf {\bibinfo {volume} {337}},\ \bibinfo {pages} {1069} (\bibinfo {year} {2012})}\BibitemShut {NoStop}%
\bibitem [{\citenamefont {Zeiher}\ \emph {et~al.}(2021)\citenamefont {Zeiher}, \citenamefont {Wolf}, \citenamefont {Isaacs}, \citenamefont {Kohler},\ and\ \citenamefont {{Stamper-Kurn}}}]{zeiherTrackingEvaporativeCooling2021}%
  \BibitemOpen
  \bibfield  {author} {\bibinfo {author} {\bibfnamefont {J.}~\bibnamefont {Zeiher}}, \bibinfo {author} {\bibfnamefont {J.}~\bibnamefont {Wolf}}, \bibinfo {author} {\bibfnamefont {J.~A.}\ \bibnamefont {Isaacs}}, \bibinfo {author} {\bibfnamefont {J.}~\bibnamefont {Kohler}},\ and\ \bibinfo {author} {\bibfnamefont {D.~M.}\ \bibnamefont {{Stamper-Kurn}}},\ }\bibfield  {title} {\bibinfo {title} {Tracking {{Evaporative Cooling}} of a {{Mesoscopic Atomic Quantum Gas}} in {{Real Time}}},\ }\href {https://doi.org/10.1103/PhysRevX.11.041017} {\bibfield  {journal} {\bibinfo  {journal} {Physical Review X}\ }\textbf {\bibinfo {volume} {11}},\ \bibinfo {pages} {041017} (\bibinfo {year} {2021})}\BibitemShut {NoStop}%
\bibitem [{\citenamefont {Gorman}\ \emph {et~al.}(2018)\citenamefont {Gorman}, \citenamefont {Hemmerling}, \citenamefont {Megidish}, \citenamefont {Moeller}, \citenamefont {Schindler}, \citenamefont {Sarovar},\ and\ \citenamefont {Haeffner}}]{gormanEngineeringVibrationallyAssisted2018}%
  \BibitemOpen
  \bibfield  {author} {\bibinfo {author} {\bibfnamefont {D.~J.}\ \bibnamefont {Gorman}}, \bibinfo {author} {\bibfnamefont {B.}~\bibnamefont {Hemmerling}}, \bibinfo {author} {\bibfnamefont {E.}~\bibnamefont {Megidish}}, \bibinfo {author} {\bibfnamefont {S.~A.}\ \bibnamefont {Moeller}}, \bibinfo {author} {\bibfnamefont {P.}~\bibnamefont {Schindler}}, \bibinfo {author} {\bibfnamefont {M.}~\bibnamefont {Sarovar}},\ and\ \bibinfo {author} {\bibfnamefont {H.}~\bibnamefont {Haeffner}},\ }\bibfield  {title} {\bibinfo {title} {Engineering {{Vibrationally Assisted Energy Transfer}} in a {{Trapped-Ion Quantum Simulator}}},\ }\href {https://doi.org/10.1103/PhysRevX.8.011038} {\bibfield  {journal} {\bibinfo  {journal} {Physical Review X}\ }\textbf {\bibinfo {volume} {8}},\ \bibinfo {pages} {011038} (\bibinfo {year} {2018})}\BibitemShut {NoStop}%
\bibitem [{\citenamefont {Guo}\ \emph {et~al.}(2024)\citenamefont {Guo}, \citenamefont {Yao}, \citenamefont {Ramanjanappa}, \citenamefont {Dhar}, \citenamefont {Horvath}, \citenamefont {Pizzino}, \citenamefont {Giamarchi}, \citenamefont {Landini},\ and\ \citenamefont {N{\"a}gerl}}]{guoObservation2D1D2024}%
  \BibitemOpen
  \bibfield  {author} {\bibinfo {author} {\bibfnamefont {Y.}~\bibnamefont {Guo}}, \bibinfo {author} {\bibfnamefont {H.}~\bibnamefont {Yao}}, \bibinfo {author} {\bibfnamefont {S.}~\bibnamefont {Ramanjanappa}}, \bibinfo {author} {\bibfnamefont {S.}~\bibnamefont {Dhar}}, \bibinfo {author} {\bibfnamefont {M.}~\bibnamefont {Horvath}}, \bibinfo {author} {\bibfnamefont {L.}~\bibnamefont {Pizzino}}, \bibinfo {author} {\bibfnamefont {T.}~\bibnamefont {Giamarchi}}, \bibinfo {author} {\bibfnamefont {M.}~\bibnamefont {Landini}},\ and\ \bibinfo {author} {\bibfnamefont {H.-C.}\ \bibnamefont {N{\"a}gerl}},\ }\bibfield  {title} {\bibinfo {title} {Observation of the {{2D}}--{{1D}} crossover in strongly interacting ultracold bosons},\ }\href {https://doi.org/10.1038/s41567-024-02459-3} {\bibfield  {journal} {\bibinfo  {journal} {Nature Physics}\ }\textbf {\bibinfo {volume} {20}},\ \bibinfo {pages} {934} (\bibinfo {year} {2024})}\BibitemShut {NoStop}%
\bibitem [{\citenamefont {{Safavi-Naini}}\ \emph {et~al.}(2018)\citenamefont {{Safavi-Naini}}, \citenamefont {{Lewis-Swan}}, \citenamefont {Bohnet}, \citenamefont {G{\"a}rttner}, \citenamefont {Gilmore}, \citenamefont {Jordan}, \citenamefont {Cohn}, \citenamefont {Freericks}, \citenamefont {Rey},\ and\ \citenamefont {Bollinger}}]{safavi-nainiVerificationManyIonSimulator2018}%
  \BibitemOpen
  \bibfield  {author} {\bibinfo {author} {\bibfnamefont {A.}~\bibnamefont {{Safavi-Naini}}}, \bibinfo {author} {\bibfnamefont {R.~J.}\ \bibnamefont {{Lewis-Swan}}}, \bibinfo {author} {\bibfnamefont {J.~G.}\ \bibnamefont {Bohnet}}, \bibinfo {author} {\bibfnamefont {M.}~\bibnamefont {G{\"a}rttner}}, \bibinfo {author} {\bibfnamefont {K.~A.}\ \bibnamefont {Gilmore}}, \bibinfo {author} {\bibfnamefont {J.~E.}\ \bibnamefont {Jordan}}, \bibinfo {author} {\bibfnamefont {J.}~\bibnamefont {Cohn}}, \bibinfo {author} {\bibfnamefont {J.~K.}\ \bibnamefont {Freericks}}, \bibinfo {author} {\bibfnamefont {A.~M.}\ \bibnamefont {Rey}},\ and\ \bibinfo {author} {\bibfnamefont {J.~J.}\ \bibnamefont {Bollinger}},\ }\bibfield  {title} {\bibinfo {title} {Verification of a {{Many-Ion Simulator}} of the {{Dicke Model Through Slow Quenches}} across a {{Phase Transition}}},\ }\href {https://doi.org/10.1103/PhysRevLett.121.040503} {\bibfield  {journal} {\bibinfo  {journal} {Physical Review Letters}\ }\textbf {\bibinfo {volume} {121}},\
  \bibinfo {pages} {040503} (\bibinfo {year} {2018})}\BibitemShut {NoStop}%
\bibitem [{\citenamefont {Kaufman}\ \emph {et~al.}(2016)\citenamefont {Kaufman}, \citenamefont {Tai}, \citenamefont {Lukin}, \citenamefont {Rispoli}, \citenamefont {Schittko}, \citenamefont {Preiss},\ and\ \citenamefont {Greiner}}]{kaufmanQuantumThermalizationEntanglement2016}%
  \BibitemOpen
  \bibfield  {author} {\bibinfo {author} {\bibfnamefont {A.~M.}\ \bibnamefont {Kaufman}}, \bibinfo {author} {\bibfnamefont {M.~E.}\ \bibnamefont {Tai}}, \bibinfo {author} {\bibfnamefont {A.}~\bibnamefont {Lukin}}, \bibinfo {author} {\bibfnamefont {M.}~\bibnamefont {Rispoli}}, \bibinfo {author} {\bibfnamefont {R.}~\bibnamefont {Schittko}}, \bibinfo {author} {\bibfnamefont {P.~M.}\ \bibnamefont {Preiss}},\ and\ \bibinfo {author} {\bibfnamefont {M.}~\bibnamefont {Greiner}},\ }\bibfield  {title} {\bibinfo {title} {Quantum thermalization through entanglement in an isolated many-body system},\ }\href {https://doi.org/10.1126/science.aaf6725} {\bibfield  {journal} {\bibinfo  {journal} {Science}\ }\textbf {\bibinfo {volume} {353}},\ \bibinfo {pages} {794} (\bibinfo {year} {2016})}\BibitemShut {NoStop}%
\bibitem [{\citenamefont {Kaufman}\ and\ \citenamefont {Ni}(2021)}]{kaufmanQuantumScienceOptical2021}%
  \BibitemOpen
  \bibfield  {author} {\bibinfo {author} {\bibfnamefont {A.~M.}\ \bibnamefont {Kaufman}}\ and\ \bibinfo {author} {\bibfnamefont {K.-K.}\ \bibnamefont {Ni}},\ }\bibfield  {title} {\bibinfo {title} {Quantum science with optical tweezer arrays of ultracold atoms and molecules},\ }\href {https://doi.org/10.1038/s41567-021-01357-2} {\bibfield  {journal} {\bibinfo  {journal} {Nature Physics}\ }\textbf {\bibinfo {volume} {17}},\ \bibinfo {pages} {1324} (\bibinfo {year} {2021})}\BibitemShut {NoStop}%
\bibitem [{\citenamefont {Browaeys}\ and\ \citenamefont {Lahaye}(2020)}]{browaeysManybodyPhysicsIndividually2020}%
  \BibitemOpen
  \bibfield  {author} {\bibinfo {author} {\bibfnamefont {A.}~\bibnamefont {Browaeys}}\ and\ \bibinfo {author} {\bibfnamefont {T.}~\bibnamefont {Lahaye}},\ }\bibfield  {title} {\bibinfo {title} {Many-body physics with individually controlled {{Rydberg}} atoms},\ }\href {https://doi.org/10.1038/s41567-019-0733-z} {\bibfield  {journal} {\bibinfo  {journal} {Nature Physics}\ }\textbf {\bibinfo {volume} {16}},\ \bibinfo {pages} {132} (\bibinfo {year} {2020})}\BibitemShut {NoStop}%
\bibitem [{\citenamefont {Ritsch}\ \emph {et~al.}(2013)\citenamefont {Ritsch}, \citenamefont {Domokos}, \citenamefont {Brennecke},\ and\ \citenamefont {Esslinger}}]{ritschColdAtomsCavitygenerated2013}%
  \BibitemOpen
  \bibfield  {author} {\bibinfo {author} {\bibfnamefont {H.}~\bibnamefont {Ritsch}}, \bibinfo {author} {\bibfnamefont {P.}~\bibnamefont {Domokos}}, \bibinfo {author} {\bibfnamefont {F.}~\bibnamefont {Brennecke}},\ and\ \bibinfo {author} {\bibfnamefont {T.}~\bibnamefont {Esslinger}},\ }\bibfield  {title} {\bibinfo {title} {Cold atoms in cavity-generated dynamical optical potentials},\ }\href {https://doi.org/10.1103/RevModPhys.85.553} {\bibfield  {journal} {\bibinfo  {journal} {Reviews of Modern Physics}\ }\textbf {\bibinfo {volume} {85}},\ \bibinfo {pages} {553} (\bibinfo {year} {2013})}\BibitemShut {NoStop}%
\bibitem [{\citenamefont {Mivehvar}\ \emph {et~al.}(2021)\citenamefont {Mivehvar}, \citenamefont {Piazza}, \citenamefont {Donner},\ and\ \citenamefont {Ritsch}}]{mivehvarCavityQEDQuantum2021}%
  \BibitemOpen
  \bibfield  {author} {\bibinfo {author} {\bibfnamefont {F.}~\bibnamefont {Mivehvar}}, \bibinfo {author} {\bibfnamefont {F.}~\bibnamefont {Piazza}}, \bibinfo {author} {\bibfnamefont {T.}~\bibnamefont {Donner}},\ and\ \bibinfo {author} {\bibfnamefont {H.}~\bibnamefont {Ritsch}},\ }\bibfield  {title} {\bibinfo {title} {Cavity {{QED}} with quantum gases: New paradigms in many-body physics},\ }\href {https://doi.org/10.1080/00018732.2021.1969727} {\bibfield  {journal} {\bibinfo  {journal} {Advances in Physics}\ }\textbf {\bibinfo {volume} {70}},\ \bibinfo {pages} {1} (\bibinfo {year} {2021})}\BibitemShut {NoStop}%
\bibitem [{\citenamefont {Kirton}\ \emph {et~al.}(2019)\citenamefont {Kirton}, \citenamefont {Roses}, \citenamefont {Keeling},\ and\ \citenamefont {Dalla~Torre}}]{kirtonIntroductionDickeModel2019}%
  \BibitemOpen
  \bibfield  {author} {\bibinfo {author} {\bibfnamefont {P.}~\bibnamefont {Kirton}}, \bibinfo {author} {\bibfnamefont {M.~M.}\ \bibnamefont {Roses}}, \bibinfo {author} {\bibfnamefont {J.}~\bibnamefont {Keeling}},\ and\ \bibinfo {author} {\bibfnamefont {E.~G.}\ \bibnamefont {Dalla~Torre}},\ }\bibfield  {title} {\bibinfo {title} {Introduction to the {{Dicke Model}}: {{From Equilibrium}} to {{Nonequilibrium}}, and {{{\emph{Vice Versa}}}}},\ }\href {https://doi.org/10.1002/qute.201800043} {\bibfield  {journal} {\bibinfo  {journal} {Advanced Quantum Technologies}\ }\textbf {\bibinfo {volume} {2}},\ \bibinfo {pages} {1800043} (\bibinfo {year} {2019})}\BibitemShut {NoStop}%
\bibitem [{\citenamefont {Baumann}\ \emph {et~al.}(2010)\citenamefont {Baumann}, \citenamefont {Guerlin}, \citenamefont {Brennecke},\ and\ \citenamefont {Esslinger}}]{baumannDickeQuantumPhase2010}%
  \BibitemOpen
  \bibfield  {author} {\bibinfo {author} {\bibfnamefont {K.}~\bibnamefont {Baumann}}, \bibinfo {author} {\bibfnamefont {C.}~\bibnamefont {Guerlin}}, \bibinfo {author} {\bibfnamefont {F.}~\bibnamefont {Brennecke}},\ and\ \bibinfo {author} {\bibfnamefont {T.}~\bibnamefont {Esslinger}},\ }\bibfield  {title} {\bibinfo {title} {Dicke quantum phase transition with a superfluid gas in an optical cavity},\ }\href {https://doi.org/10.1038/nature09009} {\bibfield  {journal} {\bibinfo  {journal} {Nature}\ }\textbf {\bibinfo {volume} {464}},\ \bibinfo {pages} {1301} (\bibinfo {year} {2010})}\BibitemShut {NoStop}%
\bibitem [{\citenamefont {L{\'e}onard}\ \emph {et~al.}(2017)\citenamefont {L{\'e}onard}, \citenamefont {Morales}, \citenamefont {Zupancic}, \citenamefont {Esslinger},\ and\ \citenamefont {Donner}}]{leonardSupersolidFormationQuantum2017}%
  \BibitemOpen
  \bibfield  {author} {\bibinfo {author} {\bibfnamefont {J.}~\bibnamefont {L{\'e}onard}}, \bibinfo {author} {\bibfnamefont {A.}~\bibnamefont {Morales}}, \bibinfo {author} {\bibfnamefont {P.}~\bibnamefont {Zupancic}}, \bibinfo {author} {\bibfnamefont {T.}~\bibnamefont {Esslinger}},\ and\ \bibinfo {author} {\bibfnamefont {T.}~\bibnamefont {Donner}},\ }\bibfield  {title} {\bibinfo {title} {Supersolid formation in a quantum gas breaking a continuous translational symmetry},\ }\href {https://doi.org/10.1038/nature21067} {\bibfield  {journal} {\bibinfo  {journal} {Nature}\ }\textbf {\bibinfo {volume} {543}},\ \bibinfo {pages} {87} (\bibinfo {year} {2017})}\BibitemShut {NoStop}%
\bibitem [{\citenamefont {Morales}\ \emph {et~al.}(2018)\citenamefont {Morales}, \citenamefont {Zupancic}, \citenamefont {L{\'e}onard}, \citenamefont {Esslinger},\ and\ \citenamefont {Donner}}]{moralesCouplingTwoOrder2018}%
  \BibitemOpen
  \bibfield  {author} {\bibinfo {author} {\bibfnamefont {A.}~\bibnamefont {Morales}}, \bibinfo {author} {\bibfnamefont {P.}~\bibnamefont {Zupancic}}, \bibinfo {author} {\bibfnamefont {J.}~\bibnamefont {L{\'e}onard}}, \bibinfo {author} {\bibfnamefont {T.}~\bibnamefont {Esslinger}},\ and\ \bibinfo {author} {\bibfnamefont {T.}~\bibnamefont {Donner}},\ }\bibfield  {title} {\bibinfo {title} {Coupling two order parameters in a quantum gas},\ }\href {https://doi.org/10.1038/s41563-018-0118-1} {\bibfield  {journal} {\bibinfo  {journal} {Nature Materials}\ }\textbf {\bibinfo {volume} {17}},\ \bibinfo {pages} {686} (\bibinfo {year} {2018})}\BibitemShut {NoStop}%
\bibitem [{\citenamefont {Kongkhambut}\ \emph {et~al.}(2022)\citenamefont {Kongkhambut}, \citenamefont {Skulte}, \citenamefont {Mathey}, \citenamefont {Cosme}, \citenamefont {Hemmerich},\ and\ \citenamefont {Ke{\ss}ler}}]{kongkhambutObservationContinuousTime2022}%
  \BibitemOpen
  \bibfield  {author} {\bibinfo {author} {\bibfnamefont {P.}~\bibnamefont {Kongkhambut}}, \bibinfo {author} {\bibfnamefont {J.}~\bibnamefont {Skulte}}, \bibinfo {author} {\bibfnamefont {L.}~\bibnamefont {Mathey}}, \bibinfo {author} {\bibfnamefont {J.~G.}\ \bibnamefont {Cosme}}, \bibinfo {author} {\bibfnamefont {A.}~\bibnamefont {Hemmerich}},\ and\ \bibinfo {author} {\bibfnamefont {H.}~\bibnamefont {Ke{\ss}ler}},\ }\bibfield  {title} {\bibinfo {title} {Observation of a continuous time crystal},\ }\href {https://doi.org/10.1126/science.abo3382} {\bibfield  {journal} {\bibinfo  {journal} {Science}\ }\textbf {\bibinfo {volume} {377}},\ \bibinfo {pages} {670} (\bibinfo {year} {2022})}\BibitemShut {NoStop}%
\bibitem [{\citenamefont {Zhiqiang}\ \emph {et~al.}(2017)\citenamefont {Zhiqiang}, \citenamefont {Lee}, \citenamefont {Kumar}, \citenamefont {Arnold}, \citenamefont {Masson}, \citenamefont {Parkins},\ and\ \citenamefont {Barrett}}]{zhiqiangNonequilibriumPhaseTransition2017}%
  \BibitemOpen
  \bibfield  {author} {\bibinfo {author} {\bibfnamefont {Z.}~\bibnamefont {Zhiqiang}}, \bibinfo {author} {\bibfnamefont {C.~H.}\ \bibnamefont {Lee}}, \bibinfo {author} {\bibfnamefont {R.}~\bibnamefont {Kumar}}, \bibinfo {author} {\bibfnamefont {K.~J.}\ \bibnamefont {Arnold}}, \bibinfo {author} {\bibfnamefont {S.~J.}\ \bibnamefont {Masson}}, \bibinfo {author} {\bibfnamefont {A.~S.}\ \bibnamefont {Parkins}},\ and\ \bibinfo {author} {\bibfnamefont {M.~D.}\ \bibnamefont {Barrett}},\ }\bibfield  {title} {\bibinfo {title} {Nonequilibrium phase transition in a spin-1 {{Dicke}} model},\ }\href {https://doi.org/10.1364/OPTICA.4.000424} {\bibfield  {journal} {\bibinfo  {journal} {Optica}\ }\textbf {\bibinfo {volume} {4}},\ \bibinfo {pages} {424} (\bibinfo {year} {2017})}\BibitemShut {NoStop}%
\bibitem [{\citenamefont {Black}\ \emph {et~al.}(2003)\citenamefont {Black}, \citenamefont {Chan},\ and\ \citenamefont {Vuleti{\'c}}}]{blackObservationCollectiveFriction2003}%
  \BibitemOpen
  \bibfield  {author} {\bibinfo {author} {\bibfnamefont {A.~T.}\ \bibnamefont {Black}}, \bibinfo {author} {\bibfnamefont {H.~W.}\ \bibnamefont {Chan}},\ and\ \bibinfo {author} {\bibfnamefont {V.}~\bibnamefont {Vuleti{\'c}}},\ }\bibfield  {title} {\bibinfo {title} {Observation of {{Collective Friction Forces}} due to {{Spatial Self-Organization}} of {{Atoms}}: {{From Rayleigh}} to {{Bragg Scattering}}},\ }\href {https://doi.org/10.1103/PhysRevLett.91.203001} {\bibfield  {journal} {\bibinfo  {journal} {Physical Review Letters}\ }\textbf {\bibinfo {volume} {91}},\ \bibinfo {pages} {203001} (\bibinfo {year} {2003})}\BibitemShut {NoStop}%
\bibitem [{\citenamefont {Deist}\ \emph {et~al.}(2022{\natexlab{a}})\citenamefont {Deist}, \citenamefont {Lu}, \citenamefont {Ho}, \citenamefont {Pasha}, \citenamefont {Zeiher}, \citenamefont {Yan},\ and\ \citenamefont {{Stamper-Kurn}}}]{deistMidCircuitCavityMeasurement2022}%
  \BibitemOpen
  \bibfield  {author} {\bibinfo {author} {\bibfnamefont {E.}~\bibnamefont {Deist}}, \bibinfo {author} {\bibfnamefont {Y.-H.}\ \bibnamefont {Lu}}, \bibinfo {author} {\bibfnamefont {J.}~\bibnamefont {Ho}}, \bibinfo {author} {\bibfnamefont {M.~K.}\ \bibnamefont {Pasha}}, \bibinfo {author} {\bibfnamefont {J.}~\bibnamefont {Zeiher}}, \bibinfo {author} {\bibfnamefont {Z.}~\bibnamefont {Yan}},\ and\ \bibinfo {author} {\bibfnamefont {D.~M.}\ \bibnamefont {{Stamper-Kurn}}},\ }\bibfield  {title} {\bibinfo {title} {Mid-{{Circuit Cavity Measurement}} in a {{Neutral Atom Array}}},\ }\href {https://doi.org/10.1103/PhysRevLett.129.203602} {\bibfield  {journal} {\bibinfo  {journal} {Physical Review Letters}\ }\textbf {\bibinfo {volume} {129}},\ \bibinfo {pages} {203602} (\bibinfo {year} {2022}{\natexlab{a}})}\BibitemShut {NoStop}%
\bibitem [{\citenamefont {Deist}\ \emph {et~al.}(2022{\natexlab{b}})\citenamefont {Deist}, \citenamefont {Gerber}, \citenamefont {Lu}, \citenamefont {Zeiher},\ and\ \citenamefont {{Stamper-Kurn}}}]{deistSuperresolutionMicroscopyOptical2022}%
  \BibitemOpen
  \bibfield  {author} {\bibinfo {author} {\bibfnamefont {E.}~\bibnamefont {Deist}}, \bibinfo {author} {\bibfnamefont {J.~A.}\ \bibnamefont {Gerber}}, \bibinfo {author} {\bibfnamefont {Y.-H.}\ \bibnamefont {Lu}}, \bibinfo {author} {\bibfnamefont {J.}~\bibnamefont {Zeiher}},\ and\ \bibinfo {author} {\bibfnamefont {D.~M.}\ \bibnamefont {{Stamper-Kurn}}},\ }\bibfield  {title} {\bibinfo {title} {Superresolution {{Microscopy}} of {{Optical Fields Using Tweezer-Trapped Single Atoms}}},\ }\href {https://doi.org/10.1103/PhysRevLett.128.083201} {\bibfield  {journal} {\bibinfo  {journal} {Physical Review Letters}\ }\textbf {\bibinfo {volume} {128}},\ \bibinfo {pages} {083201} (\bibinfo {year} {2022}{\natexlab{b}})}\BibitemShut {NoStop}%
\bibitem [{\citenamefont {Kimble}(1998)}]{kimbleStrongInteractionsSingle1998}%
  \BibitemOpen
  \bibfield  {author} {\bibinfo {author} {\bibfnamefont {H.~J.}\ \bibnamefont {Kimble}},\ }\bibfield  {title} {\bibinfo {title} {Strong {{Interactions}} of {{Single Atoms}} and {{Photons}} in {{CavityQED}}},\ }\href {https://doi.org/10.1238/Physica.Topical.076a00127} {\bibfield  {journal} {\bibinfo  {journal} {Physica Scripta}\ }\textbf {\bibinfo {volume} {T76}},\ \bibinfo {pages} {127} (\bibinfo {year} {1998})}\BibitemShut {NoStop}%
\bibitem [{\citenamefont {Yan}\ \emph {et~al.}(2023)\citenamefont {Yan}, \citenamefont {Ho}, \citenamefont {Lu}, \citenamefont {Masson}, \citenamefont {{Asenjo-Garcia}},\ and\ \citenamefont {{Stamper-Kurn}}}]{yanSuperradiantSubradiantCavity2023a}%
  \BibitemOpen
  \bibfield  {author} {\bibinfo {author} {\bibfnamefont {Z.}~\bibnamefont {Yan}}, \bibinfo {author} {\bibfnamefont {J.}~\bibnamefont {Ho}}, \bibinfo {author} {\bibfnamefont {Y.-H.}\ \bibnamefont {Lu}}, \bibinfo {author} {\bibfnamefont {S.~J.}\ \bibnamefont {Masson}}, \bibinfo {author} {\bibfnamefont {A.}~\bibnamefont {{Asenjo-Garcia}}},\ and\ \bibinfo {author} {\bibfnamefont {D.~M.}\ \bibnamefont {{Stamper-Kurn}}},\ }\bibfield  {title} {\bibinfo {title} {Superradiant and {{Subradiant Cavity Scattering}} by {{Atom Arrays}}},\ }\href {https://doi.org/10.1103/PhysRevLett.131.253603} {\bibfield  {journal} {\bibinfo  {journal} {Physical Review Letters}\ }\textbf {\bibinfo {volume} {131}},\ \bibinfo {pages} {253603} (\bibinfo {year} {2023})}\BibitemShut {NoStop}%
\bibitem [{\citenamefont {Reimann}\ \emph {et~al.}(2015)\citenamefont {Reimann}, \citenamefont {Alt}, \citenamefont {Kampschulte}, \citenamefont {Macha}, \citenamefont {Ratschbacher}, \citenamefont {Thau}, \citenamefont {Yoon},\ and\ \citenamefont {Meschede}}]{reimannCavityModifiedCollectiveRayleigh2015}%
  \BibitemOpen
  \bibfield  {author} {\bibinfo {author} {\bibfnamefont {R.}~\bibnamefont {Reimann}}, \bibinfo {author} {\bibfnamefont {W.}~\bibnamefont {Alt}}, \bibinfo {author} {\bibfnamefont {T.}~\bibnamefont {Kampschulte}}, \bibinfo {author} {\bibfnamefont {T.}~\bibnamefont {Macha}}, \bibinfo {author} {\bibfnamefont {L.}~\bibnamefont {Ratschbacher}}, \bibinfo {author} {\bibfnamefont {N.}~\bibnamefont {Thau}}, \bibinfo {author} {\bibfnamefont {S.}~\bibnamefont {Yoon}},\ and\ \bibinfo {author} {\bibfnamefont {D.}~\bibnamefont {Meschede}},\ }\bibfield  {title} {\bibinfo {title} {Cavity-{{Modified Collective Rayleigh Scattering}} of {{Two Atoms}}},\ }\href {https://doi.org/10.1103/PhysRevLett.114.023601} {\bibfield  {journal} {\bibinfo  {journal} {Physical Review Letters}\ }\textbf {\bibinfo {volume} {114}},\ \bibinfo {pages} {023601} (\bibinfo {year} {2015})}\BibitemShut {NoStop}%
\bibitem [{\citenamefont {Begley}\ \emph {et~al.}(2016)\citenamefont {Begley}, \citenamefont {Vogt}, \citenamefont {Gulati}, \citenamefont {Takahashi},\ and\ \citenamefont {Keller}}]{begleyOptimizedMultiIonCavity2016}%
  \BibitemOpen
  \bibfield  {author} {\bibinfo {author} {\bibfnamefont {S.}~\bibnamefont {Begley}}, \bibinfo {author} {\bibfnamefont {M.}~\bibnamefont {Vogt}}, \bibinfo {author} {\bibfnamefont {G.~K.}\ \bibnamefont {Gulati}}, \bibinfo {author} {\bibfnamefont {H.}~\bibnamefont {Takahashi}},\ and\ \bibinfo {author} {\bibfnamefont {M.}~\bibnamefont {Keller}},\ }\bibfield  {title} {\bibinfo {title} {Optimized {{Multi-Ion Cavity Coupling}}},\ }\href {https://doi.org/10.1103/PhysRevLett.116.223001} {\bibfield  {journal} {\bibinfo  {journal} {Physical Review Letters}\ }\textbf {\bibinfo {volume} {116}},\ \bibinfo {pages} {223001} (\bibinfo {year} {2016})}\BibitemShut {NoStop}%
\bibitem [{\citenamefont {Neuzner}\ \emph {et~al.}(2016)\citenamefont {Neuzner}, \citenamefont {K{\"o}rber}, \citenamefont {Morin}, \citenamefont {Ritter},\ and\ \citenamefont {Rempe}}]{neuznerInterferenceDynamicsLight2016}%
  \BibitemOpen
  \bibfield  {author} {\bibinfo {author} {\bibfnamefont {A.}~\bibnamefont {Neuzner}}, \bibinfo {author} {\bibfnamefont {M.}~\bibnamefont {K{\"o}rber}}, \bibinfo {author} {\bibfnamefont {O.}~\bibnamefont {Morin}}, \bibinfo {author} {\bibfnamefont {S.}~\bibnamefont {Ritter}},\ and\ \bibinfo {author} {\bibfnamefont {G.}~\bibnamefont {Rempe}},\ }\bibfield  {title} {\bibinfo {title} {Interference and dynamics of light from a distance-controlled atom pair in an optical cavity},\ }\href {https://doi.org/10.1038/nphoton.2016.19} {\bibfield  {journal} {\bibinfo  {journal} {Nature Photonics}\ }\textbf {\bibinfo {volume} {10}},\ \bibinfo {pages} {303} (\bibinfo {year} {2016})}\BibitemShut {NoStop}%
\bibitem [{\citenamefont {Sch\"utz}\ \emph {et~al.}(2015)\citenamefont {Sch\"utz}, \citenamefont {J\"ager},\ and\ \citenamefont {Morigi}}]{Schuetz2015}%
  \BibitemOpen
  \bibfield  {author} {\bibinfo {author} {\bibfnamefont {S.}~\bibnamefont {Sch\"utz}}, \bibinfo {author} {\bibfnamefont {S.~B.}\ \bibnamefont {J\"ager}},\ and\ \bibinfo {author} {\bibfnamefont {G.}~\bibnamefont {Morigi}},\ }\bibfield  {title} {\bibinfo {title} {Thermodynamics and dynamics of atomic self-organization in an optical cavity},\ }\href {https://doi.org/10.1103/PhysRevA.92.063808} {\bibfield  {journal} {\bibinfo  {journal} {Phys. Rev. A}\ }\textbf {\bibinfo {volume} {92}},\ \bibinfo {pages} {063808} (\bibinfo {year} {2015})}\BibitemShut {NoStop}%
\bibitem [{\citenamefont {Arnold}\ \emph {et~al.}(2012)\citenamefont {Arnold}, \citenamefont {Baden},\ and\ \citenamefont {Barrett}}]{arnoldSelfOrganizationThresholdScaling2012}%
  \BibitemOpen
  \bibfield  {author} {\bibinfo {author} {\bibfnamefont {K.~J.}\ \bibnamefont {Arnold}}, \bibinfo {author} {\bibfnamefont {M.~P.}\ \bibnamefont {Baden}},\ and\ \bibinfo {author} {\bibfnamefont {M.~D.}\ \bibnamefont {Barrett}},\ }\bibfield  {title} {\bibinfo {title} {Self-{{Organization Threshold Scaling}} for {{Thermal Atoms Coupled}} to a {{Cavity}}},\ }\href {https://doi.org/10.1103/PhysRevLett.109.153002} {\bibfield  {journal} {\bibinfo  {journal} {Physical Review Letters}\ }\textbf {\bibinfo {volume} {109}},\ \bibinfo {pages} {153002} (\bibinfo {year} {2012})}\BibitemShut {NoStop}%
\bibitem [{\citenamefont {Niedenzu}\ \emph {et~al.}(2011)\citenamefont {Niedenzu}, \citenamefont {Grie{\ss}er},\ and\ \citenamefont {Ritsch}}]{niedenzuKineticTheoryCavity2011}%
  \BibitemOpen
  \bibfield  {author} {\bibinfo {author} {\bibfnamefont {W.}~\bibnamefont {Niedenzu}}, \bibinfo {author} {\bibfnamefont {T.}~\bibnamefont {Grie{\ss}er}},\ and\ \bibinfo {author} {\bibfnamefont {H.}~\bibnamefont {Ritsch}},\ }\bibfield  {title} {\bibinfo {title} {Kinetic theory of cavity cooling and self-organisation of a cold gas},\ }\href {https://doi.org/10.1209/0295-5075/96/43001} {\bibfield  {journal} {\bibinfo  {journal} {EPL (Europhysics Letters)}\ }\textbf {\bibinfo {volume} {96}},\ \bibinfo {pages} {43001} (\bibinfo {year} {2011})}\BibitemShut {NoStop}%
\bibitem [{\citenamefont {Landau}(1965)}]{landauTHEORYPHASETRANSITIONS1965}%
  \BibitemOpen
  \bibfield  {author} {\bibinfo {author} {\bibfnamefont {L.}~\bibnamefont {Landau}},\ }\bibfield  {title} {\bibinfo {title} {{{ON THE THEORY OF PHASE TRANSITIONS}}},\ }in\ \href {https://doi.org/10.1016/B978-0-08-010586-4.50034-1} {\emph {\bibinfo {booktitle} {Collected {{Papers}} of {{L}}.{{D}}. {{Landau}}}}}\ (\bibinfo  {publisher} {Elsevier},\ \bibinfo {year} {1965})\ pp.\ \bibinfo {pages} {193--216}\BibitemShut {NoStop}%
\bibitem [{\citenamefont {Zhang}\ \emph {et~al.}(2021)\citenamefont {Zhang}, \citenamefont {Chen}, \citenamefont {Wu}, \citenamefont {Wang}, \citenamefont {Fan}, \citenamefont {Deng},\ and\ \citenamefont {Wu}}]{zhangObservationSuperradiantQuantum2021}%
  \BibitemOpen
  \bibfield  {author} {\bibinfo {author} {\bibfnamefont {X.}~\bibnamefont {Zhang}}, \bibinfo {author} {\bibfnamefont {Y.}~\bibnamefont {Chen}}, \bibinfo {author} {\bibfnamefont {Z.}~\bibnamefont {Wu}}, \bibinfo {author} {\bibfnamefont {J.}~\bibnamefont {Wang}}, \bibinfo {author} {\bibfnamefont {J.}~\bibnamefont {Fan}}, \bibinfo {author} {\bibfnamefont {S.}~\bibnamefont {Deng}},\ and\ \bibinfo {author} {\bibfnamefont {H.}~\bibnamefont {Wu}},\ }\bibfield  {title} {\bibinfo {title} {Observation of a superradiant quantum phase transition in an intracavity degenerate {{Fermi}} gas},\ }\href {https://doi.org/10.1126/science.abd4385} {\bibfield  {journal} {\bibinfo  {journal} {Science}\ }\textbf {\bibinfo {volume} {373}},\ \bibinfo {pages} {1359} (\bibinfo {year} {2021})}\BibitemShut {NoStop}%
\bibitem [{\citenamefont {N{\'e}el}(1953)}]{neelThermoremanentMagnetizationFine1953}%
  \BibitemOpen
  \bibfield  {author} {\bibinfo {author} {\bibfnamefont {L.}~\bibnamefont {N{\'e}el}},\ }\bibfield  {title} {\bibinfo {title} {Thermoremanent {{Magnetization}} of {{Fine Powders}}},\ }\href {https://doi.org/10.1103/RevModPhys.25.293} {\bibfield  {journal} {\bibinfo  {journal} {Reviews of Modern Physics}\ }\textbf {\bibinfo {volume} {25}},\ \bibinfo {pages} {293} (\bibinfo {year} {1953})}\BibitemShut {NoStop}%
\bibitem [{\citenamefont {Knobel}\ \emph {et~al.}(2008)\citenamefont {Knobel}, \citenamefont {Nunes}, \citenamefont {Socolovsky}, \citenamefont {De~Biasi}, \citenamefont {Vargas},\ and\ \citenamefont {Denardin}}]{knobelSuperparamagnetismOtherMagnetic2008}%
  \BibitemOpen
  \bibfield  {author} {\bibinfo {author} {\bibfnamefont {M.}~\bibnamefont {Knobel}}, \bibinfo {author} {\bibfnamefont {W.~C.}\ \bibnamefont {Nunes}}, \bibinfo {author} {\bibfnamefont {L.~M.}\ \bibnamefont {Socolovsky}}, \bibinfo {author} {\bibfnamefont {E.}~\bibnamefont {De~Biasi}}, \bibinfo {author} {\bibfnamefont {J.~M.}\ \bibnamefont {Vargas}},\ and\ \bibinfo {author} {\bibfnamefont {J.~C.}\ \bibnamefont {Denardin}},\ }\bibfield  {title} {\bibinfo {title} {Superparamagnetism and {{Other Magnetic Features}} in {{Granular Materials}}: {{A Review}} on {{Ideal}} and {{Real Systems}}},\ }\href {https://doi.org/10.1166/jnn.2008.15348} {\bibfield  {journal} {\bibinfo  {journal} {Journal of Nanoscience and Nanotechnology}\ }\textbf {\bibinfo {volume} {8}},\ \bibinfo {pages} {2836} (\bibinfo {year} {2008})}\BibitemShut {NoStop}%
\bibitem [{\citenamefont {Mottl}\ \emph {et~al.}(2012)\citenamefont {Mottl}, \citenamefont {Brennecke}, \citenamefont {Baumann}, \citenamefont {Landig}, \citenamefont {Donner},\ and\ \citenamefont {Esslinger}}]{mottlRotonTypeModeSoftening2012}%
  \BibitemOpen
  \bibfield  {author} {\bibinfo {author} {\bibfnamefont {R.}~\bibnamefont {Mottl}}, \bibinfo {author} {\bibfnamefont {F.}~\bibnamefont {Brennecke}}, \bibinfo {author} {\bibfnamefont {K.}~\bibnamefont {Baumann}}, \bibinfo {author} {\bibfnamefont {R.}~\bibnamefont {Landig}}, \bibinfo {author} {\bibfnamefont {T.}~\bibnamefont {Donner}},\ and\ \bibinfo {author} {\bibfnamefont {T.}~\bibnamefont {Esslinger}},\ }\bibfield  {title} {\bibinfo {title} {Roton-{{Type Mode Softening}} in a {{Quantum Gas}} with {{Cavity-Mediated Long-Range Interactions}}},\ }\href {https://doi.org/10.1126/science.1220314} {\bibfield  {journal} {\bibinfo  {journal} {Science}\ }\textbf {\bibinfo {volume} {336}},\ \bibinfo {pages} {1570} (\bibinfo {year} {2012})}\BibitemShut {NoStop}%
\bibitem [{\citenamefont {Helson}\ \emph {et~al.}(2023)\citenamefont {Helson}, \citenamefont {Zwettler}, \citenamefont {Mivehvar}, \citenamefont {Colella}, \citenamefont {Roux}, \citenamefont {Konishi}, \citenamefont {Ritsch},\ and\ \citenamefont {Brantut}}]{helsonDensitywaveOrderingUnitary2023}%
  \BibitemOpen
  \bibfield  {author} {\bibinfo {author} {\bibfnamefont {V.}~\bibnamefont {Helson}}, \bibinfo {author} {\bibfnamefont {T.}~\bibnamefont {Zwettler}}, \bibinfo {author} {\bibfnamefont {F.}~\bibnamefont {Mivehvar}}, \bibinfo {author} {\bibfnamefont {E.}~\bibnamefont {Colella}}, \bibinfo {author} {\bibfnamefont {K.}~\bibnamefont {Roux}}, \bibinfo {author} {\bibfnamefont {H.}~\bibnamefont {Konishi}}, \bibinfo {author} {\bibfnamefont {H.}~\bibnamefont {Ritsch}},\ and\ \bibinfo {author} {\bibfnamefont {J.-P.}\ \bibnamefont {Brantut}},\ }\bibfield  {title} {\bibinfo {title} {Density-wave ordering in a unitary {{Fermi}} gas with photon-mediated interactions},\ }\href {https://doi.org/10.1038/s41586-023-06018-3} {\bibfield  {journal} {\bibinfo  {journal} {Nature}\ }\textbf {\bibinfo {volume} {618}},\ \bibinfo {pages} {716} (\bibinfo {year} {2023})}\BibitemShut {NoStop}%
\bibitem [{\citenamefont {Vukics}\ \emph {et~al.}(2019)\citenamefont {Vukics}, \citenamefont {Dombi}, \citenamefont {Fink},\ and\ \citenamefont {Domokos}}]{vukicsFinitesizeScalingPhotonblockade2019}%
  \BibitemOpen
  \bibfield  {author} {\bibinfo {author} {\bibfnamefont {A.}~\bibnamefont {Vukics}}, \bibinfo {author} {\bibfnamefont {A.}~\bibnamefont {Dombi}}, \bibinfo {author} {\bibfnamefont {J.~M.}\ \bibnamefont {Fink}},\ and\ \bibinfo {author} {\bibfnamefont {P.}~\bibnamefont {Domokos}},\ }\bibfield  {title} {\bibinfo {title} {Finite-size scaling of the photon-blockade breakdown dissipative quantum phase transition},\ }\href {https://doi.org/10.22331/q-2019-06-03-150} {\bibfield  {journal} {\bibinfo  {journal} {Quantum}\ }\textbf {\bibinfo {volume} {3}},\ \bibinfo {pages} {150} (\bibinfo {year} {2019})}\BibitemShut {NoStop}%
\bibitem [{\citenamefont {Di~Terlizzi}\ \emph {et~al.}(2024)\citenamefont {Di~Terlizzi}, \citenamefont {Gironella}, \citenamefont {{Herraez-Aguilar}}, \citenamefont {Betz}, \citenamefont {Monroy}, \citenamefont {Baiesi},\ and\ \citenamefont {Ritort}}]{diterlizziVarianceSumRule2024}%
  \BibitemOpen
  \bibfield  {author} {\bibinfo {author} {\bibfnamefont {I.}~\bibnamefont {Di~Terlizzi}}, \bibinfo {author} {\bibfnamefont {M.}~\bibnamefont {Gironella}}, \bibinfo {author} {\bibfnamefont {D.}~\bibnamefont {{Herraez-Aguilar}}}, \bibinfo {author} {\bibfnamefont {T.}~\bibnamefont {Betz}}, \bibinfo {author} {\bibfnamefont {F.}~\bibnamefont {Monroy}}, \bibinfo {author} {\bibfnamefont {M.}~\bibnamefont {Baiesi}},\ and\ \bibinfo {author} {\bibfnamefont {F.}~\bibnamefont {Ritort}},\ }\bibfield  {title} {\bibinfo {title} {Variance sum rule for entropy production},\ }\href {https://doi.org/10.1126/science.adh1823} {\bibfield  {journal} {\bibinfo  {journal} {Science}\ }\textbf {\bibinfo {volume} {383}},\ \bibinfo {pages} {971} (\bibinfo {year} {2024})}\BibitemShut {NoStop}%
\bibitem [{\citenamefont {Su}\ \emph {et~al.}(2025)\citenamefont {Su}, \citenamefont {Douglas}, \citenamefont {Szurek}, \citenamefont {H{\'e}bert}, \citenamefont {Krahn}, \citenamefont {Groth}, \citenamefont {Phelps}, \citenamefont {Markovi{\'c}},\ and\ \citenamefont {Greiner}}]{suFastSingleAtom2025}%
  \BibitemOpen
  \bibfield  {author} {\bibinfo {author} {\bibfnamefont {L.}~\bibnamefont {Su}}, \bibinfo {author} {\bibfnamefont {A.}~\bibnamefont {Douglas}}, \bibinfo {author} {\bibfnamefont {M.}~\bibnamefont {Szurek}}, \bibinfo {author} {\bibfnamefont {A.~H.}\ \bibnamefont {H{\'e}bert}}, \bibinfo {author} {\bibfnamefont {A.}~\bibnamefont {Krahn}}, \bibinfo {author} {\bibfnamefont {R.}~\bibnamefont {Groth}}, \bibinfo {author} {\bibfnamefont {G.~A.}\ \bibnamefont {Phelps}}, \bibinfo {author} {\bibfnamefont {O.}~\bibnamefont {Markovi{\'c}}},\ and\ \bibinfo {author} {\bibfnamefont {M.}~\bibnamefont {Greiner}},\ }\bibfield  {title} {\bibinfo {title} {Fast single atom imaging for optical lattice arrays},\ }\href {https://doi.org/10.1038/s41467-025-56305-y} {\bibfield  {journal} {\bibinfo  {journal} {Nature Communications}\ }\textbf {\bibinfo {volume} {16}},\ \bibinfo {pages} {1017} (\bibinfo {year} {2025})}\BibitemShut {NoStop}%
\bibitem [{\citenamefont {Luo}\ \emph {et~al.}(2024)\citenamefont {Luo}, \citenamefont {Zhang}, \citenamefont {Koh}, \citenamefont {Wilson}, \citenamefont {Chu}, \citenamefont {Holland}, \citenamefont {Rey},\ and\ \citenamefont {Thompson}}]{luoMomentumexchangeInteractionsBragg2024}%
  \BibitemOpen
  \bibfield  {author} {\bibinfo {author} {\bibfnamefont {C.}~\bibnamefont {Luo}}, \bibinfo {author} {\bibfnamefont {H.}~\bibnamefont {Zhang}}, \bibinfo {author} {\bibfnamefont {V.~P.~W.}\ \bibnamefont {Koh}}, \bibinfo {author} {\bibfnamefont {J.~D.}\ \bibnamefont {Wilson}}, \bibinfo {author} {\bibfnamefont {A.}~\bibnamefont {Chu}}, \bibinfo {author} {\bibfnamefont {M.~J.}\ \bibnamefont {Holland}}, \bibinfo {author} {\bibfnamefont {A.~M.}\ \bibnamefont {Rey}},\ and\ \bibinfo {author} {\bibfnamefont {J.~K.}\ \bibnamefont {Thompson}},\ }\bibfield  {title} {\bibinfo {title} {Momentum-exchange interactions in a {{Bragg}} atom interferometer suppress {{Doppler}} dephasing},\ }\href {https://doi.org/10.1126/science.adi1393} {\bibfield  {journal} {\bibinfo  {journal} {Science}\ }\textbf {\bibinfo {volume} {384}},\ \bibinfo {pages} {551} (\bibinfo {year} {2024})}\BibitemShut {NoStop}%
\bibitem [{\citenamefont {Periwal}\ \emph {et~al.}(2021)\citenamefont {Periwal}, \citenamefont {Cooper}, \citenamefont {Kunkel}, \citenamefont {Wienand}, \citenamefont {Davis},\ and\ \citenamefont {{Schleier-Smith}}}]{periwalProgrammableInteractionsEmergent2021b}%
  \BibitemOpen
  \bibfield  {author} {\bibinfo {author} {\bibfnamefont {A.}~\bibnamefont {Periwal}}, \bibinfo {author} {\bibfnamefont {E.~S.}\ \bibnamefont {Cooper}}, \bibinfo {author} {\bibfnamefont {P.}~\bibnamefont {Kunkel}}, \bibinfo {author} {\bibfnamefont {J.~F.}\ \bibnamefont {Wienand}}, \bibinfo {author} {\bibfnamefont {E.~J.}\ \bibnamefont {Davis}},\ and\ \bibinfo {author} {\bibfnamefont {M.}~\bibnamefont {{Schleier-Smith}}},\ }\bibfield  {title} {\bibinfo {title} {Programmable interactions and emergent geometry in an array of atom clouds},\ }\href {https://doi.org/10.1038/s41586-021-04156-0} {\bibfield  {journal} {\bibinfo  {journal} {Nature}\ }\textbf {\bibinfo {volume} {600}},\ \bibinfo {pages} {630} (\bibinfo {year} {2021})}\BibitemShut {NoStop}%
\bibitem [{\citenamefont {Kroeze}\ \emph {et~al.}(2023)\citenamefont {Kroeze}, \citenamefont {Marsh}, \citenamefont {Schuller}, \citenamefont {Hunt}, \citenamefont {Gopalakrishnan}, \citenamefont {Keeling},\ and\ \citenamefont {Lev}}]{kroezeReplicaSymmetryBreaking2023}%
  \BibitemOpen
  \bibfield  {author} {\bibinfo {author} {\bibfnamefont {R.~M.}\ \bibnamefont {Kroeze}}, \bibinfo {author} {\bibfnamefont {B.~P.}\ \bibnamefont {Marsh}}, \bibinfo {author} {\bibfnamefont {D.~A.}\ \bibnamefont {Schuller}}, \bibinfo {author} {\bibfnamefont {H.~S.}\ \bibnamefont {Hunt}}, \bibinfo {author} {\bibfnamefont {S.}~\bibnamefont {Gopalakrishnan}}, \bibinfo {author} {\bibfnamefont {J.}~\bibnamefont {Keeling}},\ and\ \bibinfo {author} {\bibfnamefont {B.~L.}\ \bibnamefont {Lev}},\ }\href@noop {} {\bibinfo {title} {Replica symmetry breaking in a quantum-optical vector spin glass}} (\bibinfo {year} {2023}),\ \Eprint {https://arxiv.org/abs/2311.04216} {arxiv:2311.04216 [cond-mat, physics:physics, physics:quant-ph]} \BibitemShut {NoStop}%
\bibitem [{\citenamefont {Ye}\ \emph {et~al.}(2023)\citenamefont {Ye}, \citenamefont {Tian}, \citenamefont {Lin}, \citenamefont {Luo}, \citenamefont {You}, \citenamefont {Hu}, \citenamefont {Zhang}, \citenamefont {Chen},\ and\ \citenamefont {Li}}]{yeUniversalQuantumOptimization2023}%
  \BibitemOpen
  \bibfield  {author} {\bibinfo {author} {\bibfnamefont {M.}~\bibnamefont {Ye}}, \bibinfo {author} {\bibfnamefont {Y.}~\bibnamefont {Tian}}, \bibinfo {author} {\bibfnamefont {J.}~\bibnamefont {Lin}}, \bibinfo {author} {\bibfnamefont {Y.}~\bibnamefont {Luo}}, \bibinfo {author} {\bibfnamefont {J.}~\bibnamefont {You}}, \bibinfo {author} {\bibfnamefont {J.}~\bibnamefont {Hu}}, \bibinfo {author} {\bibfnamefont {W.}~\bibnamefont {Zhang}}, \bibinfo {author} {\bibfnamefont {W.}~\bibnamefont {Chen}},\ and\ \bibinfo {author} {\bibfnamefont {X.}~\bibnamefont {Li}},\ }\bibfield  {title} {\bibinfo {title} {Universal {{Quantum Optimization}} with {{Cold Atoms}} in an {{Optical Cavity}}},\ }\href {https://doi.org/10.1103/PhysRevLett.131.103601} {\bibfield  {journal} {\bibinfo  {journal} {Physical Review Letters}\ }\textbf {\bibinfo {volume} {131}},\ \bibinfo {pages} {103601} (\bibinfo {year} {2023})}\BibitemShut {NoStop}%
\bibitem [{\citenamefont {Torggler}\ \emph {et~al.}(2019)\citenamefont {Torggler}, \citenamefont {Aumann}, \citenamefont {Ritsch},\ and\ \citenamefont {Lechner}}]{torgglerQuantumNQueensSolver2019}%
  \BibitemOpen
  \bibfield  {author} {\bibinfo {author} {\bibfnamefont {V.}~\bibnamefont {Torggler}}, \bibinfo {author} {\bibfnamefont {P.}~\bibnamefont {Aumann}}, \bibinfo {author} {\bibfnamefont {H.}~\bibnamefont {Ritsch}},\ and\ \bibinfo {author} {\bibfnamefont {W.}~\bibnamefont {Lechner}},\ }\bibfield  {title} {\bibinfo {title} {A {{Quantum N-Queens Solver}}},\ }\href {https://doi.org/10.22331/q-2019-06-03-149} {\bibfield  {journal} {\bibinfo  {journal} {Quantum}\ }\textbf {\bibinfo {volume} {3}},\ \bibinfo {pages} {149} (\bibinfo {year} {2019})}\BibitemShut {NoStop}%
\bibitem [{\citenamefont {Anikeeva}\ \emph {et~al.}(2021)\citenamefont {Anikeeva}, \citenamefont {Markovi{\'c}}, \citenamefont {Borish}, \citenamefont {Hines}, \citenamefont {Rajagopal}, \citenamefont {Cooper}, \citenamefont {Periwal}, \citenamefont {{Safavi-Naeini}}, \citenamefont {Davis},\ and\ \citenamefont {{Schleier-Smith}}}]{anikeevaNumberPartitioningGrovers2021}%
  \BibitemOpen
  \bibfield  {author} {\bibinfo {author} {\bibfnamefont {G.}~\bibnamefont {Anikeeva}}, \bibinfo {author} {\bibfnamefont {O.}~\bibnamefont {Markovi{\'c}}}, \bibinfo {author} {\bibfnamefont {V.}~\bibnamefont {Borish}}, \bibinfo {author} {\bibfnamefont {J.~A.}\ \bibnamefont {Hines}}, \bibinfo {author} {\bibfnamefont {S.~V.}\ \bibnamefont {Rajagopal}}, \bibinfo {author} {\bibfnamefont {E.~S.}\ \bibnamefont {Cooper}}, \bibinfo {author} {\bibfnamefont {A.}~\bibnamefont {Periwal}}, \bibinfo {author} {\bibfnamefont {A.}~\bibnamefont {{Safavi-Naeini}}}, \bibinfo {author} {\bibfnamefont {E.~J.}\ \bibnamefont {Davis}},\ and\ \bibinfo {author} {\bibfnamefont {M.}~\bibnamefont {{Schleier-Smith}}},\ }\bibfield  {title} {\bibinfo {title} {Number {{Partitioning With Grover}}'s {{Algorithm}} in {{Central Spin Systems}}},\ }\href {https://doi.org/10.1103/PRXQuantum.2.020319} {\bibfield  {journal} {\bibinfo  {journal} {PRX Quantum}\ }\textbf {\bibinfo {volume} {2}},\ \bibinfo {pages} {020319} (\bibinfo {year}
  {2021})}\BibitemShut {NoStop}%
\bibitem [{\citenamefont {Ilias}\ \emph {et~al.}(2022)\citenamefont {Ilias}, \citenamefont {Yang}, \citenamefont {Huelga},\ and\ \citenamefont {Plenio}}]{iliasCriticalityEnhancedQuantumSensing2022}%
  \BibitemOpen
  \bibfield  {author} {\bibinfo {author} {\bibfnamefont {T.}~\bibnamefont {Ilias}}, \bibinfo {author} {\bibfnamefont {D.}~\bibnamefont {Yang}}, \bibinfo {author} {\bibfnamefont {S.~F.}\ \bibnamefont {Huelga}},\ and\ \bibinfo {author} {\bibfnamefont {M.~B.}\ \bibnamefont {Plenio}},\ }\bibfield  {title} {\bibinfo {title} {Criticality-{{Enhanced Quantum Sensing}} via {{Continuous Measurement}}},\ }\href {https://doi.org/10.1103/PRXQuantum.3.010354} {\bibfield  {journal} {\bibinfo  {journal} {PRX Quantum}\ }\textbf {\bibinfo {volume} {3}},\ \bibinfo {pages} {010354} (\bibinfo {year} {2022})}\BibitemShut {NoStop}%
\bibitem [{\citenamefont {{Fern{\'a}ndez-Lorenzo}}\ and\ \citenamefont {Porras}(2017)}]{fernandez-lorenzoQuantumSensingClose2017}%
  \BibitemOpen
  \bibfield  {author} {\bibinfo {author} {\bibfnamefont {S.}~\bibnamefont {{Fern{\'a}ndez-Lorenzo}}}\ and\ \bibinfo {author} {\bibfnamefont {D.}~\bibnamefont {Porras}},\ }\bibfield  {title} {\bibinfo {title} {Quantum sensing close to a dissipative phase transition: {{Symmetry}} breaking and criticality as metrological resources},\ }\href {https://doi.org/10.1103/PhysRevA.96.013817} {\bibfield  {journal} {\bibinfo  {journal} {Physical Review A}\ }\textbf {\bibinfo {volume} {96}},\ \bibinfo {pages} {013817} (\bibinfo {year} {2017})}\BibitemShut {NoStop}%
\bibitem [{\citenamefont {Tsang}(2013)}]{tsangQuantumTransitionedgeDetectors2013}%
  \BibitemOpen
  \bibfield  {author} {\bibinfo {author} {\bibfnamefont {M.}~\bibnamefont {Tsang}},\ }\bibfield  {title} {\bibinfo {title} {Quantum transition-edge detectors},\ }\href {https://doi.org/10.1103/PhysRevA.88.021801} {\bibfield  {journal} {\bibinfo  {journal} {Physical Review A}\ }\textbf {\bibinfo {volume} {88}},\ \bibinfo {pages} {021801} (\bibinfo {year} {2013})}\BibitemShut {NoStop}%
\bibitem [{\citenamefont {Endres}\ \emph {et~al.}(2016)\citenamefont {Endres}, \citenamefont {Bernien}, \citenamefont {Keesling}, \citenamefont {Levine}, \citenamefont {Anschuetz}, \citenamefont {Krajenbrink}, \citenamefont {Senko}, \citenamefont {Vuletic}, \citenamefont {Greiner},\ and\ \citenamefont {Lukin}}]{endresAtombyatomAssemblyDefectfree2016}%
  \BibitemOpen
  \bibfield  {author} {\bibinfo {author} {\bibfnamefont {M.}~\bibnamefont {Endres}}, \bibinfo {author} {\bibfnamefont {H.}~\bibnamefont {Bernien}}, \bibinfo {author} {\bibfnamefont {A.}~\bibnamefont {Keesling}}, \bibinfo {author} {\bibfnamefont {H.}~\bibnamefont {Levine}}, \bibinfo {author} {\bibfnamefont {E.~R.}\ \bibnamefont {Anschuetz}}, \bibinfo {author} {\bibfnamefont {A.}~\bibnamefont {Krajenbrink}}, \bibinfo {author} {\bibfnamefont {C.}~\bibnamefont {Senko}}, \bibinfo {author} {\bibfnamefont {V.}~\bibnamefont {Vuletic}}, \bibinfo {author} {\bibfnamefont {M.}~\bibnamefont {Greiner}},\ and\ \bibinfo {author} {\bibfnamefont {M.~D.}\ \bibnamefont {Lukin}},\ }\bibfield  {title} {\bibinfo {title} {Atom-by-atom assembly of defect-free one-dimensional cold atom arrays},\ }\href {https://doi.org/10.1126/science.aah3752} {\bibfield  {journal} {\bibinfo  {journal} {Science}\ }\textbf {\bibinfo {volume} {354}},\ \bibinfo {pages} {1024} (\bibinfo {year} {2016})}\BibitemShut {NoStop}%
\end{thebibliography}%
\clearpage

\onecolumngrid
\pagebreak


\section*{Supplementary material}

Here we have defined $\sigma = \sqrt{\frac{k_B T}{M \nu^2}}$, which gives the standard deviation of the atomic thermal density distribution in a harmonic tweezer. From this, we obtain an approximation for $\chi$:

$$\chi_{\text{approx}} \simeq \left.\frac{\partial \langle z \rangle_{\text{approx}}}{\partial (\delta z)}\right|_{\delta z = 0} = \frac{\int_{-\infty}^{+\infty} \frac{Nz^2}{\sigma^2} \exp\left(-\frac{N}{\sigma^2}\left(\frac{\Omega^2}{6\Omega_c^2}(k^2 z^2 - 3)z^2 + \frac{1}{2}z^2\right)\right)dz}{\int_{-\infty}^{+\infty} \exp\left(-\frac{N}{\sigma^2}\left(\frac{\Omega^2}{6\Omega_c^2}(k^2 z^2 - 3)z^2 + \frac{1}{2}z^2\right)\right)dz}$$

$$= \frac{\pi(3N(r^2-1)I_{-\frac{1}{4}}\left(\frac{3N(r^2-1)^2}{16k^2 r^2 \sigma^2}\right) - (3N(r^2-1)^2 + 8k^2 r^2 \sigma^2)I_{\frac{1}{4}}\left(\frac{3N(r^2-1)^2}{16k^2 r^2 \sigma^2}\right) + 3N(r^2-1)^2(I_{\frac{3}{4}}\left(\frac{3N(r^2-1)^2}{16k^2 r^2 \sigma^2}\right) - I_{\frac{5}{4}}\left(\frac{3N(r^2-1)^2}{16k^2 r^2 \sigma^2}\right)))}{4\sqrt{2}k^2 r^2(r^2-1)\sigma^2 K_{\frac{1}{4}}\left(\frac{3N(r^2-1)^2}{16k^2 r^2 \sigma^2}\right)},$$

(S34)

where $r \equiv \Omega/\Omega_c$ and $I_\alpha(x)$ and $K_\alpha(x)$ are modified Bessel functions of the first and second kind respectively.

We can similarly obtain an integral expression for $\chi$ where we keep the sinusoidal form of the cavity potential. In this case, the average antiphase displacement is

$$\langle z \rangle = \frac{\int_{-\infty}^{+\infty} z \exp\left(-\frac{N}{4k^2 \sigma^2}\left(-r^2 + 2k^2(z - \delta z)^2 + r^2 \cos(2kz)\right)\right)dz}{\int_{-\infty}^{+\infty} \exp\left(-\frac{N}{4k^2 \sigma^2}\left(-r^2 + 2k^2(z - \delta z)^2 + r^2 \cos(2kz)\right)\right)dz}.$$

(S35)

Thus, the expression for $\chi$ becomes

$$\chi \simeq \left.\frac{\partial \langle z \rangle}{\partial (\delta z)}\right|_{\delta z = 0} = \frac{\int_{-\infty}^{+\infty} \frac{Nz^2}{\sigma^2} \exp\left(-\frac{N}{4k^2 \sigma^2}\left(-r^2 + 2k^2 z^2 + r^2 \cos(2kz)\right)\right)dz}{\int_{-\infty}^{+\infty} \exp\left(-\frac{N}{4k^2 \sigma^2}\left(-r^2 + 2k^2 z^2 + r^2 \cos(2kz)\right)\right)dz}.$$

(S36)

By numerically computing the integrals in Eq. (S36), we obtain the theory curves shown in Fig. 4C of the main text.

---

[S1] M. Endres, H. Bernien, A. Keesling, H. Levine, E. R. Anschuetz, A. Krajenbrink, C. Senko, V. Vuletic, M. Greiner, and M. D. Lukin, Atom-by-atom assembly of defect-free one-dimensional cold atom arrays, Science **354**, 1024 (2016).
[S2] Z. Yan, J. Ho, Y.-H. Lu, S. J. Masson, A. Asenjo-Garcia, and D. M. Stamper-Kurn, Superradiant and Subradiant Cavity Scattering by Atom Arrays, Physical Review Letters **131**, 253603 (2023).
[S3] M. Aspelmeyer, T. J. Kippenberg, and F. Marquardt, Cavity optomechanics, Rev. Mod. Phys. **86**, 1391 (2014).
[S4] V. Vuletić and S. Chu, Laser cooling of atoms, ions, or molecules by coherent scattering, Phys. Rev. Lett. **84**, 3787 (2000).
[S5] V. Vuletić, H. W. Chan, and A. T. Black, Three-dimensional cavity doppler cooling and cavity sideband cooling by coherent scattering, Phys. Rev. A **64**, 033405 (2001).
[S6] P. Domokos and H. Ritsch, Mechanical effects of light in optical resonators, J. Opt. Soc. Am. B **20**, 1098 (2003).
[S7] P. Domokos and H. Ritsch, Collective cooling and self-organization of atoms in a cavity, Phys. Rev. Lett. **89**, 253003 (2002).
[S8] J. K. Asbóth, P. Domokos, H. Ritsch, and A. Vukics, Self-organization of atoms in a cavity field: Threshold, bistability, and scaling laws, Phys. Rev. A **72**, 053417 (2005).
[S9] S. Schütz, S. B. Jäger, and G. Morigi, Thermodynamics and dynamics of atomic self-organization in an optical cavity, Phys. Rev. A **92**, 063808 (2015).
[S10] S. B. Jäger, S. Schütz, and G. Morigi, Mean-field theory of atomic self-organization in optical cavities, Phys. Rev. A **94**, 023807 (2016).
[S11] M. Nairn, L. Giannelli, G. Morigi, S. Slama, B. Olmos, and S. B. Jäger, Spin-self-organization in an optical cavity facilitated by inhomogeneous broadening (2024), arXiv:2407.19706 [cond-mat.quant-gas].
[S12] M. Le Bellac, F. Mortessagne, and G. G. Batrouni, *Equilibrium and Non-Equilibrium Statistical Thermodynamics* (Cambridge University Press, 2004).
[S13] When the interatomic separation is an integer (half-integer) multiple of $\lambda$ the dominant mode is the center-of-mass (antiphase) mode where neighboring atoms move in phase (in phase opposition).


\end{document}